\documentclass[manuscript,screen]{acmart}

\usepackage{amssymb}
\usepackage{amsmath}
\usepackage{graphicx}
\usepackage{rotating}

\AtBeginDocument{%
  }

\setcopyright{acmlicensed}
\copyrightyear{2025}
\acmYear{2025}
\begin{document}

\title{TrainBo: An Interactive Robot-assisted Scenario
Training System for Older Adults with Dementia}

%
\author{FUNG Kwong Chiu}
\email{kcfungag@connect.ust.hk}
\orcid{0000-0001-8253-3641}
\affiliation{%
  \institution{Hong Kong University of Science and Technology}
  \city{Hong Kong}
  \country{CHINA}
}

\author{MOW Wai Ho}
\email{eewhmow@ust.hk}
\orcid{0000-0003-1804-0476}
\affiliation{%
  \institution{Hong Kong University of Science and Technology}
  \city{Hong Kong}
  \country{CHINA}
}
  








\begin{abstract}
Dementia is an overall decline in memory and cognitive skills severe enough to reduce an elder's ability to perform everyday activities. There is an increasing need for accessible technologies for cognitive training to slow down the cognitive decline. With the ability to provide instant feedback and assistance, social robotic systems have been proven effective in enhancing learning abilities across various age groups. This study focuses on the design of an interactive robot-assisted scenario training system (TrainBo) with self-determination theory, derives design requirements through formative and formal studies and the system's usability is also be evaluated. A pilot test is conducted on seven older adults with dementia in an elderly care center in Hong Kong for four weeks. Our finding shows that older adults with dementia have an improvement in behavioural engagement (+4.44\%), emotional engagement (+4.49\%), and intrinsic motivation (+6.45\%) after using Trainbo. These findings can provide valuable insights into the development of more captivating interactive robots for extensive training purposes.
\end{abstract}



\received{13 May 2025}

\maketitle

\section{Introduction} \label{sec:introduction}
Dementia is a neurocognitive disorder characterized by a significant decline in at least one of the cognition domains, including executive function, complex attention, language, learning, memory, perceptual-motor, or social cognition \citep{Emmady2025}. The decline reflects a persistent and progressive change from the individual's previously attained cognitive baseline and is not solely attributable to an acute episode of delirium \citep{cipriani2020daily}. People with dementia lose the ability in managing daily living, and require full-time care in the severe stage \citep{Sachdev2014}. Currently, over 50 million people in the world have dementia, the prevalence rate of dementia is expected to growth continuously with the aging population and the number is expected to increase to 131 million by 2050 \citep{Arvanitakis2019}.


Cognitive stimulation therapy (CST) is widely adopted in the management of symptoms of dementia, which consists of a variety of brain-stimulating activities, such as  memory-based activity (e.g.: memory boxes and flashcards) to stimulate memory recall \citep{Spector2003} ; language and communication activity (e.g.: word games, puzzles, and storytelling) to enhance language skills; problem-solving activity (e.g.: Sudoku, crosswords, and puzzles) to improve logic skills; and some others like social interaction, creative and artistic activities, personalization, and adaptability \citep{Valenzuela2006}. Thus, CST can improve the life quality of between older adults with dementia and  their caregivers, including mood, confidence, and social interaction.

To improve motivation, well-being and quality of life for both older adults with dementia in cognitive training, self-determination theory (SDT) is applied, which consists of three core psychological needs: autonomy, competence, and relatedness \citep{Gagne2022UAS}. 

Social robots integrated with cognitive training are becoming increasingly acknowledged as effective resources for enhancing cognitive functions in older adults with dementia \citep{Mezrar2022ASR}. Their interactive design elements enable dynamic, user-responsive interactions, enhance older adults' perceived digital relatedness by fostering a sense of belonging and social connection. These robots present a variety of activities, ranging from social interaction, puzzles to word games, catering to diverse interests and cognitive abilities. They usually offer immediate feedback to enable users and their caregivers after each activity to monitor progress and areas that need improvement. Additionally, robots tend to be more cost-effective than traditional in-person intervention activities, which often demand greater resources. 

Despite the growing interest in human-robot interaction within the field of healthcare robotics \citep{han2024human, kurazume2022development, gongor2025remarkable}, there remains a limited understanding of the specific needs and preferences of older adults, particularly those with cognitive impairments such as dementia. Existing studies fail to adequately address the unique challenges faced by this population, including sensory and visual limitations, communication difficulties, and varying levels of cognitive ability \citep{guthrie2018combined}. This oversight can lead to interventions not fully tailored to users, ultimately diminishing their effectiveness and engagement.

Additionally, the integration of psychological theories, particularly SDT, in the design of interactive robot-assisted scenario training has not been thoroughly explored. SDT highlights the significance of autonomy, competence, and relatedness in fostering motivation and engagement \citep{ryan2017self}. Tailored cognitive training that aligns with personal interests fosters autonomy by allowing choice and control; Activities integrating SDT principles (e.g., goal-setting, feedback and motivation design) improve adherence to cognitive exercises, slow decline, and make tasks engaging and achievable \citep{Gagne2022UAS}.  However, many current interventions overlook these principles, which may hinder their potential to enhance learning experiences for older adults with dementia.

Furthermore, existing robot research tends to prioritize technical aspects and functionalities over the holistic user experience \citep{coronado2022evaluating}. This narrow focus neglects the critical importance of user satisfaction and emotional engagement, both essential for any educational intervention's success. A deeper understanding of how older adults interact with these systems, emotional responses, and overall satisfaction is vital for designing more effective and appealing robot-assisted training environments.

To achieve this goal, we have developed \emph{TrainBo}, an interactive robot-assisted scenario training system aimed at engaging older adults with dementia. In order to address the existing gaps, we explore following two research questions (RQs): 

\begin{itemize}
    \item RQ1: How does the integration of self-determination theory principles into the interactive robot-assisted scenario learning system affect learning engagement and motivation of older adults with dementia?
    \item RQ2: How does the usability of interactive robot-assisted scenario learning system impact user satisfaction of older adults with dementia?
\end{itemize}

The main contributions of this work can be summarized as follows:
\begin{itemize}
    \item We have constructed design requirements for older adults with dementia by collaborating with experts (therapists) and target end-users (older adults with dementia).
    \item We propose an interactive robot-assisted scenario training system, which integrates with speech-to-text function, visual aids and human-like interactions to support the individualized training tasks.
    \item We have conducted a comprehensive evaluation and encompassed a four-week user study to exemplify the effectiveness and usability of the advanced system.
\end{itemize}

This paper is organized as follows structure.

Section \ref{sec:related_work} outlines related work of assistive technology, social robot and cognitive training systems.

Sections \ref{sec:formative_study} elaborates the formative study and how to collection the requirement from stakeholders. 

Sections \ref{sec:system_design} describes the system design and development procedures.

Section \ref{sec:formal_study} discusses the experiment procedures, user study, result and findings.

Section \ref{sec:evaluation} evaluate the system based on the data collected from the experiment.

Section \ref{sec:discussion} summarizes key insights and discusses limitations.

Section \ref{sec:conclusion} concludes this paper and suggests some future research directions.

\begin{table}[t]
\caption{Publications related to assistive tools for older adults with dementia}
\label{table:related_publication}
\begin{tabular}{llllllll}
\hline
ID& Title                                                                                                                                                                                      & Ref   & Aim                                                                                                                                                                                                                                                  & Location  & Lan     & N & Tools                                                                                      \\ \hline
1    & \begin{tabular}[c]{@{}l@{}}Socially Assistive Robots \\ and Sensory Feedback for \\ Engaging Older Adults in \\ Cognitive Activities\end{tabular}                                          & \cite{Nault2025}        & \begin{tabular}[c]{@{}l@{}}To motivate older adults to engage \\ in cognitive activities to slow down \\ cognitive decline and enhance \\ engagement through physical, \\ cognitive, and social stimuli \\ during robot-mediated tasks.\end{tabular} & UK        & English & 9 & \begin{tabular}[c]{@{}l@{}}Socially \\ assistive\\ robots\end{tabular}                     \\
2    & \begin{tabular}[c]{@{}l@{}}An Internet-of-Things \\ Solution to Assist Inde-\\ pendent Living and Social \\ Connectedness in Elderly\end{tabular}                                          & \cite{Forkan2019}        & \begin{tabular}[c]{@{}l@{}}To develop IoT-based system \\ that supports independent living \\ for older adults by monitoring \\ their daily activities, enhancing \\ safety, automating routine tasks.\end{tabular}                                  & Australia & English & 6 & \begin{tabular}[c]{@{}l@{}}Smart home\\  sensors\end{tabular}                              \\
3    & \begin{tabular}[c]{@{}l@{}}Discourse Behavior of \\ Older Adults Interacting \\ with a Dialogue Agent \\ Competent in Multiple \\ Topics\end{tabular}                                      & \cite{Razavi2022}        & \begin{tabular}[c]{@{}l@{}}To evaluate conversational \\ agents that provides realistic \\ conversational practice for \\ older adults who are at risk \\ of social isolation or social \\ anxiety.\end{tabular}                                     & USA       & English & 9 & AI agents                                                                                  \\
4    & \begin{tabular}[c]{@{}l@{}}"For me at 90, it's going to \\ be difficult": feasibility of \\ using iPad video-\\ conferencing with older \\ adults in long-term \\ aged care\end{tabular}   & \cite{moyle2020me}        & \begin{tabular}[c]{@{}l@{}}It investigates user experiences, \\ perceptions, and challenges of \\ older adults related to adopting \\ new technology for communi-\\ cation with family and friends.\end{tabular}                                     & Australia & English & 6 & \begin{tabular}[c]{@{}l@{}}iPads and \\ video-\\ conferencing\\ softwares\end{tabular}     \\
5    & \begin{tabular}[c]{@{}l@{}}"It's Our Gang" - Promoting \\ Social Inclusion for People \\ with Dementia by Using \\ Digital Communication \\ Support in a Group Activity\end{tabular}       & \cite{Samuelsson2020}        & \begin{tabular}[c]{@{}l@{}}To understand how older adults \\ with dementia experience using \\ a web-based communication \\ support application during group \\ activities, focus on their \\ perception of social inclusion.\end{tabular}           & Sweden    & English & 5 & \begin{tabular}[c]{@{}l@{}}Tablets with \\ digital \\ communi-\\ cation tools\end{tabular} \\
6    & \begin{tabular}[c]{@{}l@{}}TrainBo: Designing an \\ Interactive Robot-assisted \\ scenario-based Training \\ System for Older Adults \\ with Dementia: A \\ Preliminary Study\end{tabular} & Our work & \begin{tabular}[c]{@{}l@{}}To design an interactive robot-\\ assisted scenario training system \\ for older adults with dementia, \\ that improves their behavioural \\ engagement, emotion engagem-\\ ent and intrinsic motivation.\end{tabular}    & Hong Kong & Chinese & 7 & \begin{tabular}[c]{@{}l@{}}Socially \\ assistive\\  robots\end{tabular}                    \\ \hline
\end{tabular}
\end{table}

\section{Related Work} \label{sec:related_work}

\subsection{Assistive Tools for Older Adults with Dementia}
The integration of digital cognitive stimulation, interactive games, video communication, and VR technologies offers promising, multifaceted approaches to support cognitive health and quality of life in older adults with dementia.
Recent efforts have explored digital adaptations of cognitive stimulation. Digital stimulation has been shown to have the benefit in cognition as traditional in-person cognitive training sessions in a randomized controlled trial \cite{Tsantali2017Cognitive}. In the systematic review of common digital technologies for older adults with dementia, indicating that new technologies have potential benefits in improving quality of life in terms of reducing loneliness and social isolation \cite{Rai2022Digital, ryu2020simple}. 
For interactive and game-based approaches, a nostalgic interactive somatosensory game for older adults with MCI was developed and discovered that game was easier to use than traditional methods, improving willingness and motivation for rehabilitation \cite{Chang2022Interactive}. Another brain stimulation game with CST also developed to reduce the risk of dementia, the analysis showed that 19 out of 30 participants demonstrated improvement in cognitive ability within seven days of training \cite{Udjaja2021Healthy,Lobbia2018Cognitive}. 
For communication enhancements via video-conferencing, research showed that video-conferencing was feasible to improve communication for those with dementia in long-term care \citep{moyle2020me}.
For Virtual Reality (VR) technology, VR applications are popular in the healthcare area. A VR-based neuro-psychological testing system to diagnose MCI. The system effectively combined digital cognitive parameters with Electroencephalography signals to diagnose MCI, which was promising for early detection of dementia \cite{Xue2023VRNPT}. For the research of using VR to promote the well-being of older adults with dementia, result found that VR positively affected the emotional, social, and functional aspects \cite{Appel2021VR}. Although numerous technologies have been developed to support older adults with dementia, there is limited research on the usability and accessibility challenges of these new technologies as these older adults with dementia face difficulties due to cognitive decline, sensory impairments, or lack of familiarity with technology. This can affect consistent use and engagement.

\subsection{Social Robots for Older Adults with Dementia}
Social assistive robots show potential to improve emotional well-being, social interaction, and cognitive engagement for older adults with dementia, especially when designed with user preferences and sensory feedback in mind. The PARO social robot designed in care settings for older adults with dementia, the work highlighted that PARO could reduce negative emotions and improve social engagement \cite{Hung2019Benefits}, the research also align with the robot therapy that can significantly reduced agitation and increased social engagement in older adults with dementia \cite{Yu2022Socially}. Robot-assisted cognitive games can be a valuable addition to existing cognitive stimulation activities, improving usability and acceptability by catering to older adults' preferences and unique needs \cite{Gasteiger2021Robot}. And the integration of socially assistive robots and sensory feedback could enhance engagement of older adults in cognitive activities with participatory design workshops, prototyping, and multimodal sensory feedback mechanisms \cite{Nault2025Socially}. However, another research of Telenoid humanoid robots as a communication tool for older adults with dementia, highlighed the technical difficulties and the need for operator training in the application of robots in healthcare \cite{Moyle2022Therapeutic}.

\subsection{Robot-assisted Cognitive Training}
Robot-assisted cognitive training emphasizes how robots can help people to learn in practical, real-world situations. They support humans in gaining skills, knowledge, or behaviours through interactive scenarios that reflect daily life or particular challenges. Pilot study on scenario-based training with colour-print images and VR images, memory tasks are assigned to older adults with dementia, results showed that both groups demonstrated positive training effects in the home setting and convenience shop \cite{man2012evaluation}. It bridged robotics with geriatric care, offering a potential tool to support ageing-in-place and improving quality of life. Another digital game that includes exercises consisting of shopping in a hypermarket, they associated with episodic declarative memory, naming, calculation, and organization, discovered that episodic memory training with exercises is effective in improving memory skills, thus providing healthier aging and daily function \cite{moran2024serious, banducci2017active}. However, 
effective use of these equipments requires overcoming technical issues and providing adequate operator training, which can limit widespread adoption in care settings, which affect user experience and consistent engagement.

Table \ref{table:related_publication} shows related works on assistive tools for older adults with dementia.

\section{Formative Study} \label{sec:formative_study}

\begin{figure}
     \centering
         \centering
         \includegraphics[width=\textwidth]{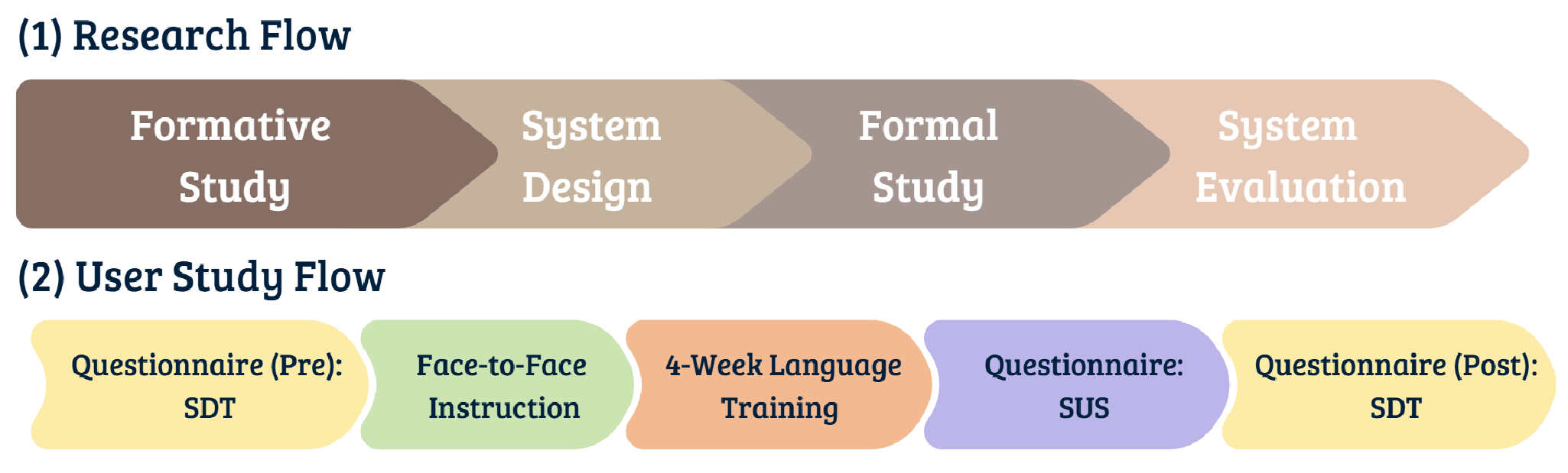}
         \caption{research flow and user study flow.}
          \label{fig:research flow and user study flow}
\end{figure}

To understand the needs of older adults with dementia when interacting with the robot-assisted scenario training system, we conducted semi-structured interviews with older adults, Figure \ref{fig:research flow and user study flow} illustrates the research flow and user study flow. 

\subsection{Participants}

\paragraph{Regulations}
To participate in this study, older adults were required to meet the following criteria:
(1) be diagnosed with dementia; 
(2) be able to speak Cantonese; and 
(3) have no medical or physical disabilities that could affect their interaction with robots. In addition, all older adults were experienced with the use of tablets. 

We obtained informed consent from the guardian prior to starting the experiment. Participation was completely voluntary and contingent on consent. The study is reviewed by the The University Institutional Review Board (IRB). 

These older adults are over 70 years old, they are diagnosed with dementia by medical practitioners, all speak the Cantonese language; they live in Hong Kong for long periods of time and currently live in the same elderly care center, they are familiar with Chinese culture and have the experience of using smart electronic devices. 
Originally, this study included 11 older adults with dementia (E1 -- E11; $\Bar{M}=86.66$, $\sigma=8.07$; one male). However, three females (E1, E5 \& E11) and one male (E3) were sick and they withdrew from the study after the formative study. Finally, seven participants (E2, E4, E6 -- E10; $\Bar{M}=87.39$, $\sigma=10.10$; all females) participated in the experiment (formal study, training and user study).

\subsection{Survey and Interview}
This study included face-to-face surveys and interviews, aimed at exploring the preferences of older adults with dementia regarding scenario training.

\paragraph{Survey} 
We conducted a face-to-face survey, which was reviewed for clarity by an experienced therapist. Following the collection of survey responses, the first author, in collaboration with another author, performed a thematic analysis of the data \citep{braun2006using}, here are survey questions:
\begin{enumerate}
    \item When answering a question, what challenges or difficulties do you usually face? 
     \item Can you describe what makes it feel challenging for you?
    \item After completing a training task, what kind of feedback or rewards would you find most motivating or satisfying? For example, would you like visual feedback, sounds, or something else?
\end{enumerate}

\paragraph{Survey Results} 
Most respondents expressed a strong preference for visual-based instructions, with 70\% indicating a desire for content to be read aloud and for the ability to control their response time. The majority reported needing to read or listen to the questions multiple times to fully grasp the key content. All participants emphasized the importance of receiving encouragement throughout the learning process. Many older adults with dementia found existing digital training solutions to be user-unfriendly, which hindered their ability to comprehend the content.

Additionally, most respondents identified as illiterate and preferred having someone read the training tasks aloud to aid their understanding. Many faced difficulties with reading small text or text in lighter colours due to visual impairments. Nearly all participants expressed a preference for preset time limits when answering questions. Furthermore, they all indicated a desire for various forms of encouragement, such as animations and sound effects, to enhance their overall learning experience. By synthesizing the survey results, we developed more focused questions for the individual interviews.

\paragraph{Interview} 
The same group of participants was recruited using a purposive sampling method. The interviews lasted for one hour. After obtaining consent from the participants, we introduced the objectives of the study. Each interview session was recorded using the iPad's recording function. We posed various questions to explore the needs of older adults with dementia in relation to digital training tasks. Here are interview questions:
\begin{enumerate}
    \item Do you prefer text-based or visual-based instructions? Why?
    \item When you are having training, do you prefer to read the questions yourself, or be read out loud? Why?
    \item Do you need to read or listen to the questions multiple times to understand the content? Why?
    \item When answering questions, do you prefer to control the time yourself, or follow a pre-set time limit? Why?
    \item After completing a question, what forms of encouragement, such as animation and text, do you want to receive? Why?
\end{enumerate}

The first author utilized the Speech-to-Text (STT) function to transcribe all recorded content, then the thematic analysis was conducted in collaboration between authors. The first author performed the initial coding to generate preliminary codes. This was followed by two rounds of discussions to group and refine these codes, ensuring a comprehensive understanding of the feedback received.

\paragraph{Interview Results} 
Based on the interview study, we have distilled the following five design requirements (DRs).

\textbf{DR 1: Visual-Based Instructions}
Older adults with dementia expressed a strong preference for visual aids, as they facilitate understanding and engagement.
Survey results indicated that 70\% of participants desired content to be read aloud and preferred visual instructions. Many participants identified as inability to read and write due to the low level of education, they are also struggled with reading small text or text in lighter colours due to presbyopia.

\textbf{DR 2: Encouragement and Positive Feedback}
Continuous encouragement is vital for enhancing motivation and engagement among older adults with dementia.
All participants emphasized the importance of receiving encouragement throughout the learning process. They expressed a desire for various forms of encouragement, such as animations and sound effects, to enhance their overall learning experience.

\textbf{DR 3: Large and Readable Font}

A larger font size improves readability, which is critical for older adults with presbyopia.
Participants reported difficulties with small text and preferred larger fonts (font size should be at least 72 points) to aid comprehension. 

\textbf{DR 4: Familiar and Relatable Content}
Familiar scenarios help older adults relate to the training, making it easier to engage and learn.
Participants indicated a preference for training scenarios that mimic real-life situations, enhancing their ability to comprehend and apply learned skills.

\textbf{DR 5: Customizable Response Time}
Allowing users to control their response time can reduce anxiety and improve engagement.
The majority of participants preferred preset time limits for answering questions, indicating a need for flexibility in pacing.

\section{System Design}  \label{sec:system_design}
After collecting the user requirements from the formative study, we introduce the system design and details of TrainBo system.

\subsection{System Hardware} 
\begin{figure}
\centering
     \includegraphics[width=0.3\textwidth]{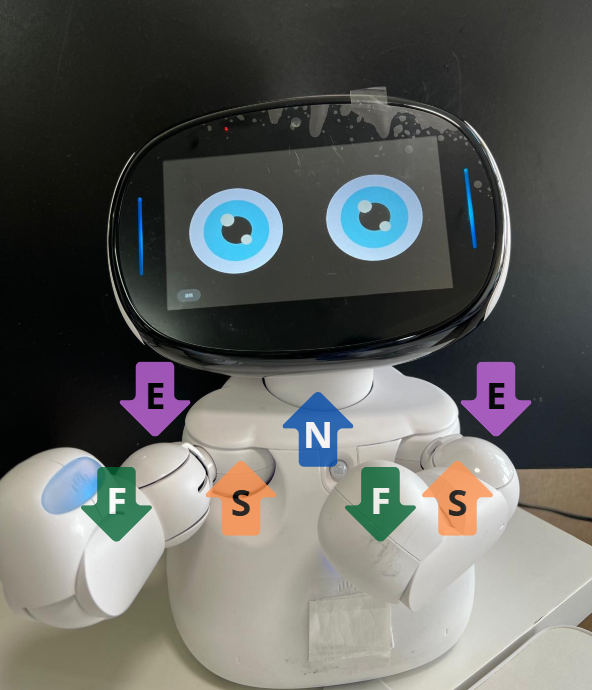}
       \caption{TrainBo has seven motors to control seven parts of its body, including a neck (N), two shoulders (S), two elbows (E) and two fists (F).}
         \label{fig:Robot-Kebbi}
\end{figure}

TrainBo is an interactive scenario training system developed based on the robot Kebbi Air, which is produced by NUWA Robotics company \citep{nuwa2024kebbi}. TrainBo features one head, two movable hands, and three wheels. As illustrated in Figure \ref{fig:Robot-Kebbi}, it is equipped with seven motors that control various parts of its body. These motors allow TrainBo to turn its head left and right, as well as move it up and down. Moreober, TrainBo can mimic some human-like hand movements, including encouragement, making fists and cheering. With its two swivel wheels and one auxiliary wheel, TrainBo can ``dance'' by rolling around. 

\subsection{System Design with Self-determination Theory} 
SDT is a psychological framework that underscores the significance of motivation and engagement by fulfilling three core supports: autonomy, competence, and relatedness \citep{ryan2017self}. In the context of older adults with dementia, SDT provides a guiding framework for designing interactive robots that enhance cognitive abilities and improve quality of life \citep{liu2023older}. By incorporating SDT principles into human-robot interactions, such as gestures, mobility, and animacy, robots can engage older adults through various interactions, such as singing, dancing, and responsive reactions \citep{moro2019}. 

\textbf{Autonomy support} the degree in which older adults perceive control over their interactions with robots \citep{pirhonen2020could}. For example, TrainBo offers customizable activities and responses, allowing older adults to make decisions that empower their engagement.

\textbf{Competence support} enhances the confidence of older adults to interact with robots and complete tasks \citep{fasola2012using}. TrainBo provides adaptive feedback and presents challenges that are appropriate for user skill levels \citep{cansev2021interactive}. For instance, positive reinforcement, such as ``You are so smart!" after successfully following a dance routine, can improve their confidence and encourage interaction.

\textbf{Relatedness support} encompasses the sense of connection and social bonding that older adults experience with TrainBo \citep{ostrowski2019older}. TrainBo simulates social interactions that alleviate feelings of loneliness and isolation by employing empathetic gestures, facial expressions, and conversational cues. \cite{paterson2023social}. For example, TrainBo strengthens the connection with older adults with dementia by presenting supportive gestures with encouraging facial expressions.

\subsection{Interaction and Motivation Design}
\begin{figure}
\centering
\includegraphics[width=0.7\linewidth]{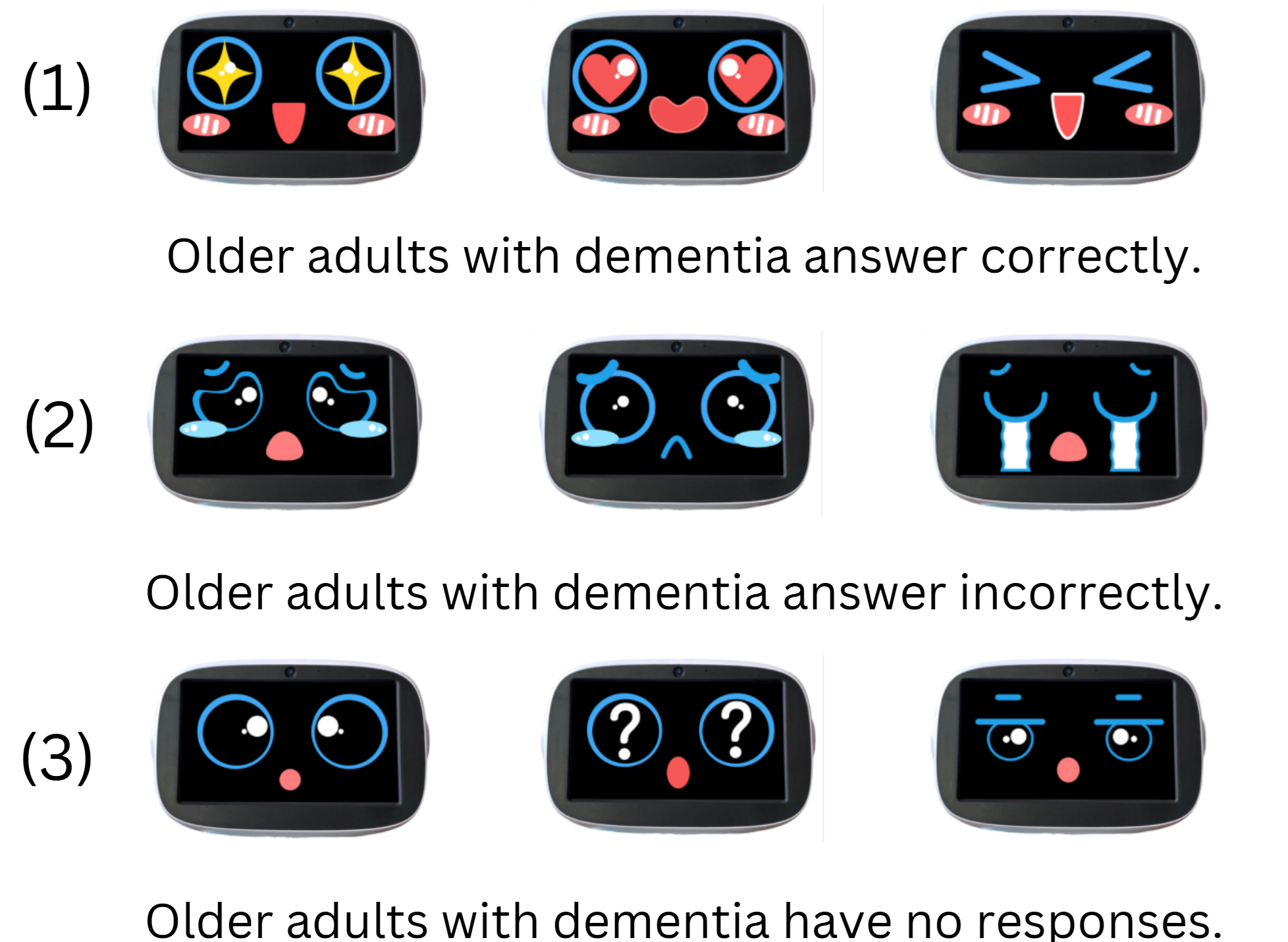}
\caption{Displayed animations when older adults with dementia (1) answer correctly, (2) answer incorrectly, and (3) have no response to the robot.}
\label{fig:response}
\end{figure}

In the interactive robot training, three human-like interaction elements are implemented to enhance learning motivation and engagement among older adults with dementia \textbf{(DR2)}. Participants can control their response speed to answer the question \textbf{(DR5)}, when answer the question correctly, Trainbo praises them by saying, ``You have done a good job!" accompanied by a positive facial expression, a nodding head, and a ``hurray" gesture (see Figure \ref{fig:response} (1)). If they answer incorrectly, Trainbo encourages them with, "You are very close! Let's try again!" while displaying a supportive facial expression, moving its head, and using an empathetic gesture (see Figure \ref{fig:response} (2)).

\begin{figure}
\centering
     \includegraphics[width=0.9\textwidth]{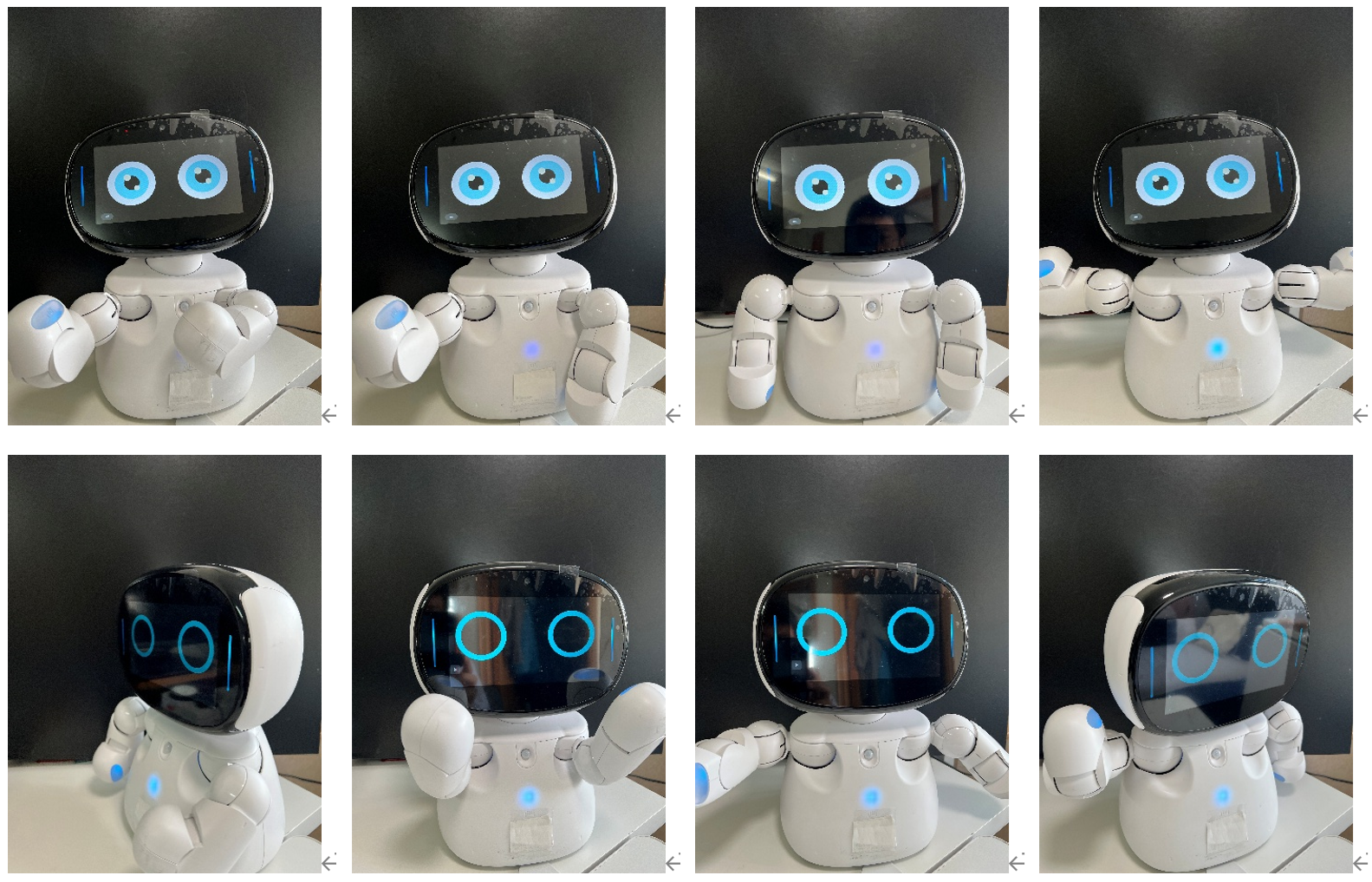}
       \caption{Top: TrainBo is singing a song; Bottom: TrainBo is dancing}
         \label{fig:dance_song}
\end{figure}

In instances where there is no interaction, TrainBo motivates the participants by asking, ``Hey! Are you thinking of the answer? No worries! I am here with you!" using a powerful facial expression, nodding its head, and performing a thinking gesture (see Figure \ref{fig:response} (3)). After completing the training, TrainBo expresses appreciation for the participants' engagement by showing an energetic face, shaking its head, and clapping, all while playing an uplifting song (see Figure \ref{fig:dance_song}).

\subsection{Personalized Design Approach}
The cognitive learning system was developed using Android Studio, incorporating automatic speech recognition (ASR) technology, which includes STT and Text-to-Speech (TTS) functions. The system can facilitate the conversion of spoken language into text and supports Traditional Chinese (Cantonese) transcription and translation. The pre-programmed system compares users input with transcription and translation, enabling the older adults to receive instant feedback and improve their language skills.

\subsection{Pedagogy Design}
\begin{figure}
\centering
\includegraphics[width=\linewidth]{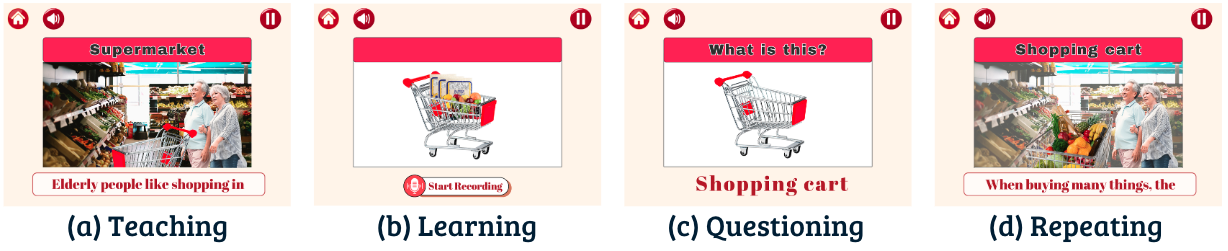}
\caption{The question samples: (a) teaching, (b) learning, (c) questionning, and (d) repeating.}
\label{fig:Question Samples}
\end{figure}

The pedagogy was co-designed with an education professional and reviewed by an experienced therapist. And make sure the font size and picture of the content are comfortable for older adults to read \textbf{(DR3)}. The application of this study consists of a four-stage training process, comprising teaching, learning, repeating, and questioning (Fig. \ref{fig:Question Samples}), which is discussed in detail as follows.

\textit{Teaching.} 
The teaching stage provides clear guidance and structure \textbf{(DR1)}, helping older adults with dementia feel confident and supported by \cite{surr2017effective}. \emph{TrainBo} introduces new concepts in an accessible way, reducing frustration and increasing willingness to engage. The teaching process captures attention by presenting information interactively and relatably, encouraging active participation through demonstrations or examples \citep{das2024chapter}. Furthermore, teaching stimulates the brain by introducing new information and concepts, building foundational knowledge essential for progressing to more complex tasks \citep{boekaerts1997self, liverman2015cognitive}.

\textit{Learning.}
The learning stage allows older adults with dementia to apply what they have been taught, fostering a sense of accomplishment \citep{bar2021fostering}. This process encourages curiosity and self-directed exploration, boosting intrinsic motivation \citep{paradowski2024predictors, puebla2022mobile}. \emph{TrainBo} keeps participants actively involved by requiring them to process and internalize information while providing opportunities for hands-on practice, making the experience more interactive. Through the learning process, older adults with dementia can strengthen memory retention and problem-solving skills through active engagement \citep{fury2024active}.

\textit{Questioning.}
The questioning stage encourages active thinking and problem-solving, which can be rewarding and motivating \citep{fury2024active}. \emph{TrainBo} provides immediate feedback, helping older adults with dementia understand their progress and areas for improvement. The questioning process keeps users mentally stimulated by challenging them to apply what they have learned, creating a dynamic and interactive experience that prevents monotony \citep{rai2020individual, goodall2021use}. Furthermore, questioning enhances critical thinking and decision-making skills, promoting deeper understanding by requiring users to analyze and synthesize information \citep{chang2020effects}.

\textit{Repeating.}
The repeating stage reinforces learning among older adults with dementia, building confidence and reducing fear of failure \citep{buchanan2011role}. \emph{TrainBo} provides a sense of mastery, encouraging users to continue engaging with the system. The repeating process creates a predictable structure, which can be comforting and motivating \citep{cipriani2013repetitive}. Additionally, repetition allows older adults with dementia to refine their skills, making the process more enjoyable and less stressful \citep{smith2011memory}. Repeating strengthens memory consolidation through repetition, which is particularly important for older adults, and improves recall and automaticity, making tasks easier to perform over time \citep{dewar2014boosting}.

\subsection{Training Scenario Design}
\emph{TrainBo} consists of three sets of scenario training, including supermarket, Chinese restaurant, and transportation. These scenarios are chosen because they are familiar to older adults with dementia in the Hong Kong culture \textbf{(DR4)}.
\emph{TrainBo} follows the following steps to train the participants in the supermarket scenario: 

\paragraph{Supermarket}
The scenario for older adults with dementia focuses on teaching essential skills for navigating a local supermarket environment, emphasizing the use of shopping carts and baskets, safety awareness, and appropriate social interactions. They can learn the functions of the shopping tools, understanding when to use a basket versus a cart, and practice safe shopping behaviours, such as pushing carts slowly and avoiding obstacles. The scenario encourages them to enhance communication skills, including how to ask for items and respond to potential rejections from family members. Additionally, \emph{TrainBo} addresses the payment process, reinforcing the importance of patience and gratitude towards cashiers. Through interactive discussions and practice, participants can build confidence and independence, ultimately enhancing their ability to engage in everyday shopping activities while fostering a sense of accomplishment and social connection.

\begin{enumerate} 
\item Introduction to supermarket shopping (e.g., The supermarket is a place where we can buy food and other items. Shopping can be a fun activity!) 
\item Identifying shopping carts and baskets (e.g., In the supermarket, we use shopping carts and baskets to carry our items.) 
\item Using a basket for few items (e.g., When buying a few items, we can use a basket.) 
\item Using a cart for many items (e.g., When buying many items, we can use a shopping cart.) 
\item Safety awareness (e.g., We should push the cart slowly and be careful not to bump into others.) 
\item Polite interactions (e.g., If we accidentally bump into someone, we should say, ``I'm sorry.") 
\item Payment process (e.g., After shopping, we need to go to the checkout to pay for our items.) 
\end{enumerate}

\paragraph{Restaurant}
The restaurant scenario aims to teach older adults with dementia appropriate behaviours and etiquette when dining out. They can learn the steps involved in entering a Chinese restaurant, including how to obtain a waiting number and the importance of following to staff's instructions. The scenario emphasizes proper dining etiquette, such as using utensils correctly. Participants can practice ordering food, thanking staff for their service, and understanding the payment process after the meal. By engaging in these activities, older adults can gain the skills needed to enjoy dining experiences, fostering social interactions and enhancing their overall dining confidence.
\emph{TrainBo} follows the following steps to train the participants in the restaurant scenario: 

\begin{enumerate}
    \item Introduction to dining out
(e.g., Many people enjoy dining at restaurants. It can be a special experience!)
    \item Entering the restaurant
(e.g., Upon arrival, we may need to wait for a table. We can take a number from the ticket machine.)
    \item Following to the staff's instructions
(e.g., We should follow carefully to the staff's instructions when it is our turn.)
    \item Ordering food
(e.g., We can look at the menu and tell our family what we would like to eat.)
    \item Dining etiquette
(e.g., We should use utensils properly.)
    \item Thanking the staff
(e.g., When our food arrives, we should thank the staff for their service.)
    \item Payment process
(e.g., After finishing our meal, we will go to the checkout to pay.)
\end{enumerate}

\paragraph{Transportation}
The transportation scenario is designed to help older adults with dementia understand and use public transportation in Hong Kong, specifically railways. They can learn about navigating train stations, including identifying ticket machines and understanding the importance of safety on platforms, such as standing behind the yellow line and waiting patiently for trains. The scenario covers appropriate behaviour inside trains, encouraging participants to maintain a quiet demeanour and respect fellow passengers. Through interactive role-playing, participants can practice the steps for boarding and exiting trains, reinforcing their ability to travel independently. 
\emph{TrainBo} follows the following steps to train the participants in the transportation scenario: 

\begin{enumerate}
    \item Introduction to railways
(e.g., The railway is a fast and convenient way to travel to different places.)
    \item Navigating the train station
(e.g., At the train station, we will find ticket machines and gates to enter.)
    \item Safety on the platform
(e.g., We should stand behind the yellow line and not run on the platform.)
    \item Boarding the train
(e.g., When the train arrives, we will wait for the doors to open before boarding.)
    \item Behaviour inside the train
(e.g., Inside the train, we can sit quietly or hold onto the handrails if standing.)
    \item Respecting other passengers
(e.g., We should keep quiet and not disturb other passengers during the ride.)
    \item Exiting the train
(e.g., When we arrive at our destination, we will wait for our turn to exit the train.)
\end{enumerate}

\section{Formal Study} \label{sec:formal_study}
This section investigates the effectiveness of the \emph{TrainBo} system and answer research questions (RQs). The study consists of training sessions and a triangulation process to ensure the reliability and validity of the evaluation, including observation, questionnaires and interview. The system is modified based on user feedback.

\begin{figure}
     \centering
         \centering
         \includegraphics[width=.6\textwidth]{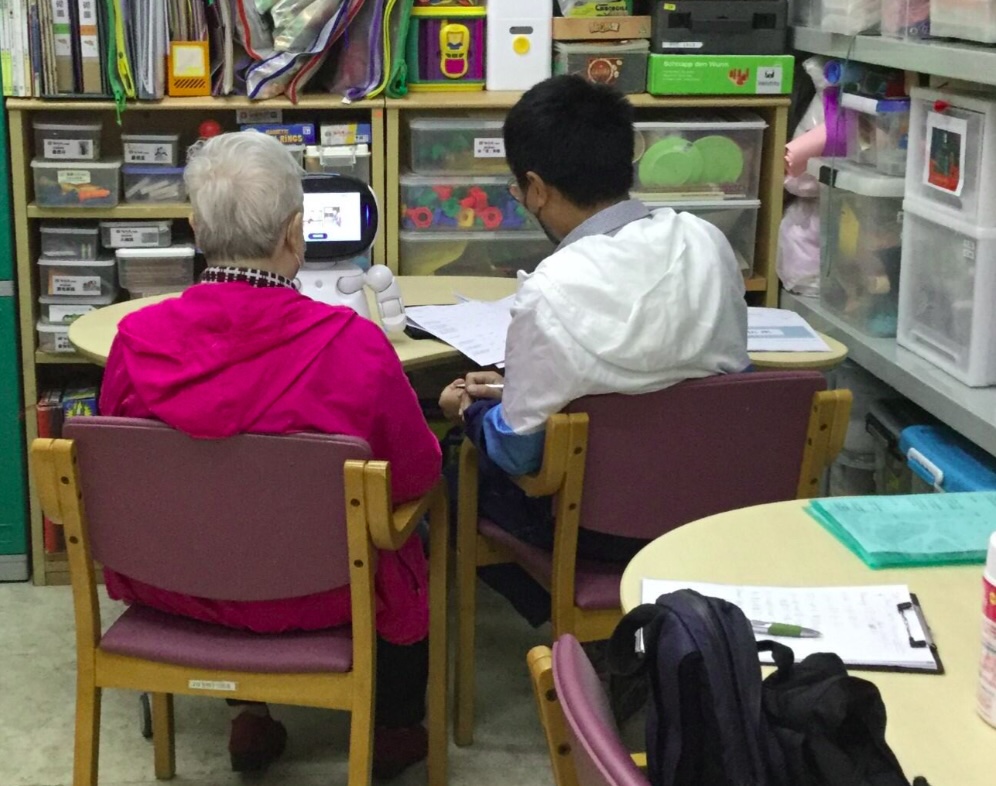}
         \caption{An instructor was supporting an older adults with dementia in an elderly centre.}
          \label{fig:elderly centre setting}
\end{figure}

As shown in Figure \ref{fig:elderly centre setting}, each older adult with dementia received technical support from an instructor.

\subsection{Training Sessions}
Each older adult with dementia participated in four sessions (i.e., one week one session) during the training. Each training session was held for 60 minutes. The instructor provided pre-training to the older adults on interacting with the robots, such as the robot's possible actions and communication procedures. After the pre-training, older adults would follow the procedure to complete the scenario training, which is outlined below:
\begin{itemize}
    \item Week 1 (Session 1):
\begin{enumerate}
    \item Older adults were required to complete a 10-minute pre-questionnaire.
    \item Older adults participated in a 60-minute scenario training about the supermarket (Scenario 1, pre-test). 
\end{enumerate}

    \item Week 2 (Session 2):
\begin{enumerate}
    \item Older adults participated in a 60-minute scenario about the restaurant (Scenario 2).
\end{enumerate}

\item Week 3 (Session 3):
\begin{enumerate}
    \item Older adults took part in a 60-minute scenario about the transportation (Scenario 3).
\end{enumerate}

\item Week 4 (Session 4):
\begin{enumerate}
    \item Older adults underwent a 60-minute scenario training about the supermarket (Scenario 1, post-test).
    \item Older adults were required to complete a 10-minute post-questionnaire.
    \item Older adults participated in a 10-minute interview.
\end{enumerate}
\end{itemize}

\subsection{Triangulation Process}

\paragraph{Observation}
During the training sessions, an experienced therapist joined us for the observation, allowing for a comprehensive analysis of the interactions between the older adults and the \emph{TrainBo} system. We observe their colour preference, understanding of equipment, patience and engagement, potential communication challenges, need of visual aids, and scenario preference.

\paragraph{Questionnaires to evaluate the SDT}
The pre-test and post-test questionnaires of SDT consisted of four variables (behaviour, emotional, cognitive engagement and intrinsic motivation) categorized into motivation and engagement. Each variable consists of three sub-questions. In total, there are 12 questions. Each variable was assessed using a 5-point Likert scale from ``Strongly Disagree" (1) to ``Strongly Agree" (5). 
\begin{itemize}

\item Pre-test Questionnaire 
 
 \begin{enumerate}
\item  	Behavioural engagement

 \begin{enumerate}
\item I put in my best effort to succeed in all the training tasks. 
\item While participating in the training tasks at the elderly center, I work as hard as I can. 
\item I actively engage in all learning activities during the training tasks at the elderly center. 
  \end{enumerate}
\item Emotional engagement 
 \begin{enumerate}
\item I feel interested when participating in the training tasks at the elderly center. 
\item I feel good while doing the training tasks at the elderly center. 
\item I find the training sessions with the therapist enjoyable. 
  \end{enumerate}

\item	Cognitive engagement

 \begin{enumerate}
\item I carefully review the training tasks to ensure I understand them correctly. 
\item I consider different strategies for completing the training tasks. 
\item I try to relate what I am learning to previous knowledge and experiences.
  \end{enumerate}

\item  Intrinsic motivation of Chinese

 \begin{enumerate}
\item I find the training sessions to be joyful. 
\item I find the training sessions to be fun. 
\item I am interested in the training tasks with the therapist.
\end{enumerate}
\end{enumerate}

\item Post-test Questionnaire 
 
 \begin{enumerate}
\item  	Behavioural engagement

 \begin{enumerate}
\item I try hard to do well in all the training tasks. \item When training with the robot, I work as hard as I can. 
\item When training with the robot, I participate in all activities.
 \end{enumerate}
 
\item Emotional engagement 
 \begin{enumerate}
\item When training with the robot, I feel interested. \item When training with the robot, I feel good. 
\item I find training with the robot to be fun.
  \end{enumerate}

\item	Cognitive engagement

 \begin{enumerate}
\item I carefully review the training tasks to ensure I understand them correctly.
\item I think about different ways to approach the training tasks. 
\item I try to connect what I am learning during training to things I have learned before.

  \end{enumerate}

\item  Intrinsic motivation of Chinese

 \begin{enumerate}
\item I find training with the robot to be joyful. 
\item I find training with the robot to be fun. 
\item I am interested in the training tasks with the robot.
\end{enumerate}
\end{enumerate}
\end{itemize}

To ensure the clarity of the questionnaire, an experienced therapist reviewed the items.

\paragraph{System Usability Scale Questionnaires}
System Usability Scale (SUS) questionnaire is used to measure the impact of user satisfaction of older adult with dementia when interacting with \emph{TrainBo} system. SUS is a widely validated tool for assessing perceived usability, particularly in technology-focused interventions for this population. Following \cite{bangor2008empirical} interpretation guidelines, 5-point likert scale was utilized, from ``Strongly Disagree" (1) to ``Strongly Agree" (5). In total, there are 10 questions with odd- (positively worded) and even-numbered statements (negatively worded). The design helps mitigate biased responses. 
 To convert the responses into a score out of 100, one point was subtracted from the score for positive statements; the score was subtracted from five for negative statements. To get a score out of 100, all scores were summed and multiplied by 2.5.
 The resulting SUS scores are within the range of 0 -- 100, and higher scores indicate a higher usability rating \citep{kortum2014relationship, lewis2018system}. In general, 68 points is acceptable \citep{trend2023measuring}.
 Here is the questionnaire of the SUS:
 \begin{enumerate}
\item I think that I would like to use this system frequently.
\item I found the system unnecessarily complex.
\item I thought the system was easy to use.
\item I think that I would need the support of a technical person to be able to use this system.
\item I found the various functions in this system were well integrated.
\item I thought there was too much inconsistency in this system.
\item I would imagine that most people would learn to use this system very quickly.
\item I found the system very cumbersome to use.
\item I felt very confident using the system.
\item I needed to learn a lot of things before I could get going with this system.
\end{enumerate}
 To ensure the clarity of the questionnaire, an experienced therapist reviewed the items. 

\paragraph{Interview}
When older adults were filling questionnaires, we conducted semi-structured interviews to collect their feedback to evaluate the system. 

\begin{figure}
     \centering
         \centering
         \includegraphics[width=\textwidth]{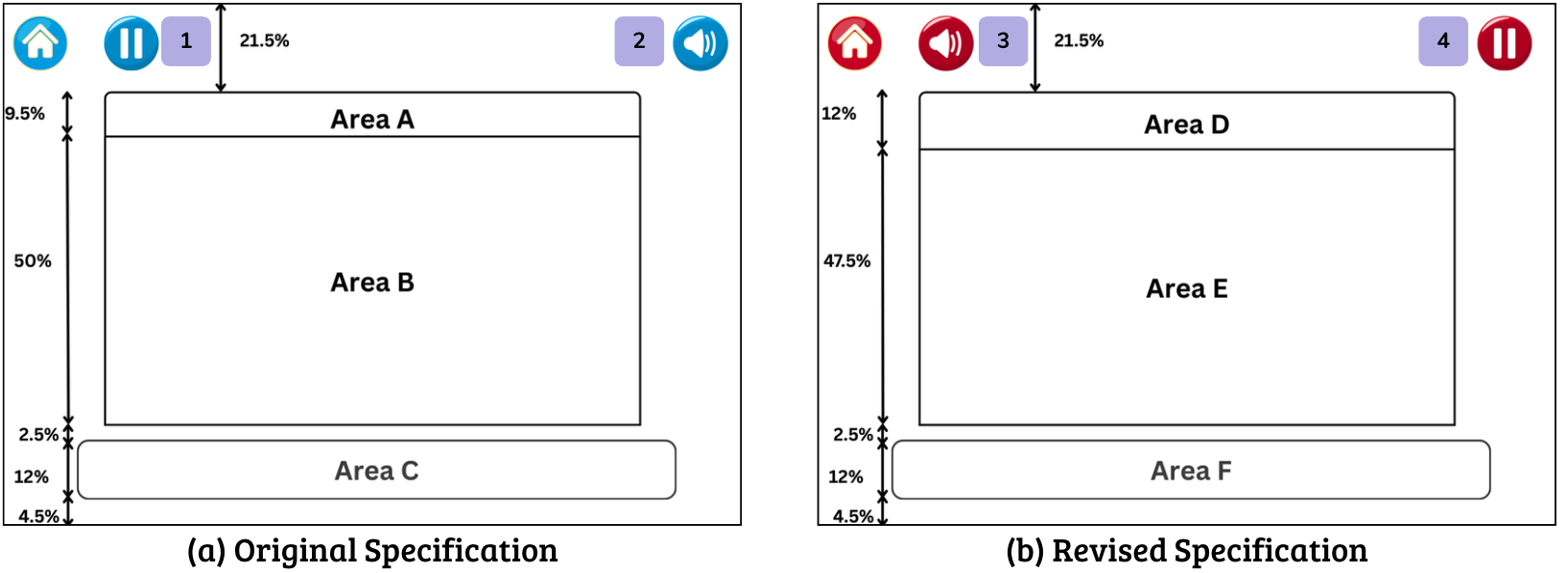}
         \caption{An overview of the specifications is as follows: (1) original and (2) revised. Areas A and D are designated for the question display, Areas B and E are for the graphic/main content display, and Areas C and F are for the answer/recording button display.}
          \label{fig:specifications}
\end{figure}

\subsection{System Modification}
\begin{figure}
     \centering
         \centering
         \includegraphics[width=\textwidth]{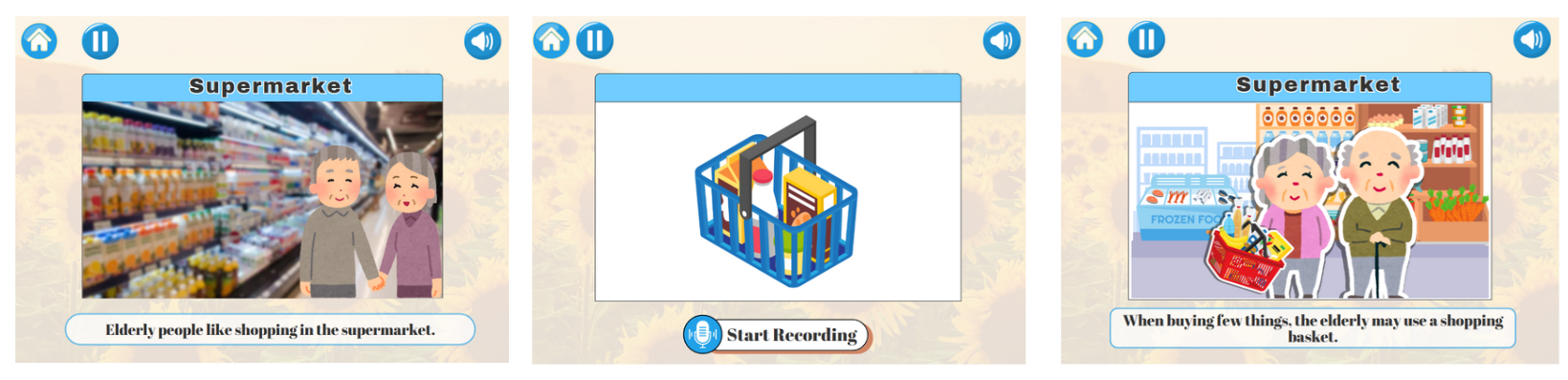}
         \caption{Interface design during the formative study}
          \label{fig:UI_formative}
\end{figure}

\begin{figure}
     \centering
         \centering
         \includegraphics[width=\textwidth]{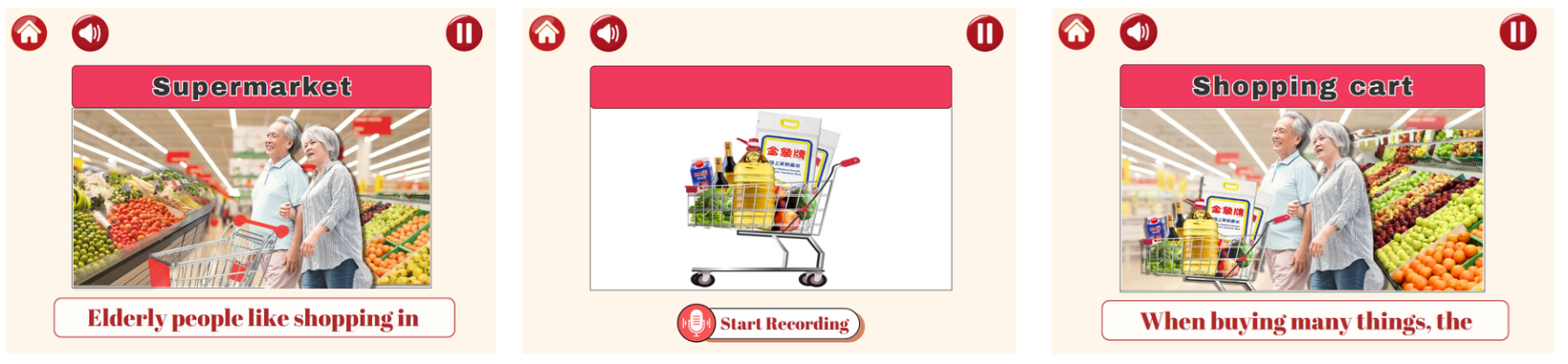}
         \caption{Revised user interface design during the formal study}
          \label{fig:UI_formal}
\end{figure}
First, we swapped the positions of the ``pause" (Figure \ref{fig:specifications}, 1 \& 4) and ``re-listen" (Figure \ref{fig:specifications}, 2 \& 3) buttons. Participants frequently pressed the ``re-listen" button with their finger, making it easier to control and locate when placed in the top right corner. From the participants' feedback, they prefer red color than blue color because red is more visible, also they consider red is associated with celebrations and prosperity in Chinese culture, therefore we changed the button colour to red \textbf{(DR1)}.

Second, participants requested a larger font size to enhance readability. Therefore, we increased 26\% of the font size in Area A , as illustrated in Figure \ref{fig:specifications}, (b), Area D. To accommodate this change, we reduced the area allocated for the graphic/main content display from 50\% (Figure \ref{fig:specifications}, (a), Area B) to 47.5\% (Figure \ref{fig:specifications}, (b), Area E) \textbf{(DR3)}.

Third, Figure \ref{fig:UI_formative} illustrated the user interface design during the formative study and Figure \ref{fig:UI_formal} illustrate the revised user interface design during the formal study, colour of all buttons were changed from blue to red to improve visibility for older adults with dementia \textbf{(DR1)}. Additionally,the background colour of the question display were changed from blue to red. font size is increased and the text colour is changed from blue to red for better readability \textbf{(DR3)}.

Furthermore, the design from cartoon images is revised to reality-oriented visuals, which serve as an effective intervention to help maintain a sense of reality and reduce confusion among participants \citep{kume2023effect} \textbf{(DR4)}. Also, participants can control the answering time by pressing ``Start Recording" buttons by themselves \textbf{(DR5)}. Regardless of whether participants answered the questions correctly, incorrectly or did not respond, \emph{TrainBo} would display different facial expressions and gestures to encourage them \textbf{(DR2)}. Additionally, \emph{TrainBo} would provide instant feedback to the participants \textbf{(DR2)}.

Finally, the answer display area was enlarged by changing from a double-line format  to a single-line format, enhancing visibility for the participants \textbf{(DR1)}.

\section{Evaluation} \label{sec:evaluation}
In this section, we analyzed the data collected from the experiment, an expert (i.e., therapist) and older adults with dementia.

\subsection{RQ1: How does the integration of SDT principles into the interactive robot-assisted scenario learning system affect learning engagement and motivation of older adults with dementia?}

\begin{figure}
     \centering
         \centering
         \includegraphics[width=.7\textwidth]{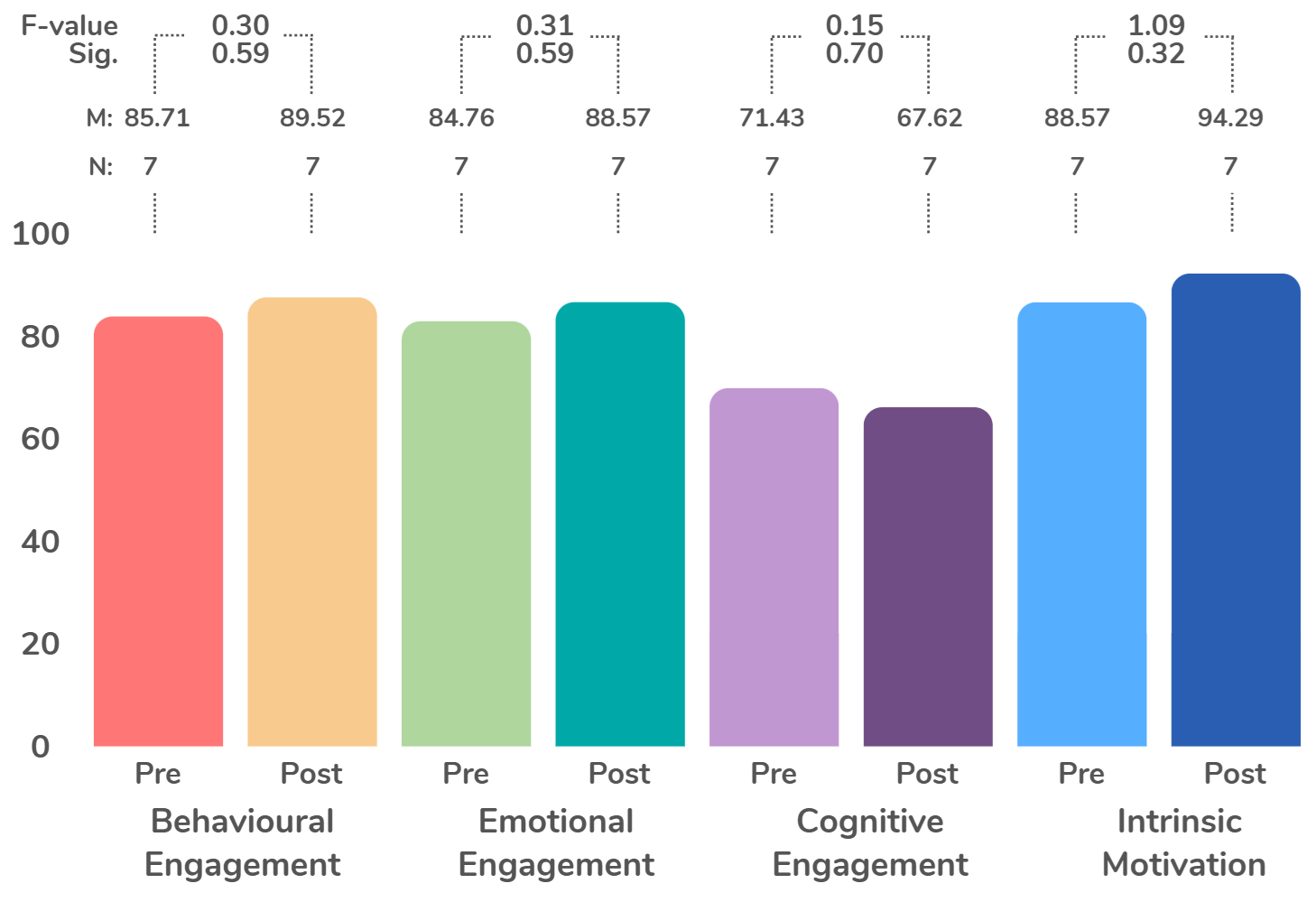}
         \caption{Summary of analysis of variance (ANOVA) for three types of engagement and one motivation.}
          \label{fig:SDT_ANOVA}
\end{figure}

Figure \ref{fig:SDT_ANOVA} depicted analysis of covariance (ANCOVA) results that they (E1–E7) exhibited improvements in three engagement and motivation compared to traditional learning methods (pre-test). The results are discussed in detail below. 

In the SDT questionnaire, older adults with dementia showed no statistically significant improvement in engagement and motivation, but there was considerable improvement.

The average score of behavioural engagement increased by 4.44\%
(pre-test: $\Bar{M}=85.71$, $\sigma=15.12$; 
post-test: $\Bar{M}=89.52$, $\sigma=10.59$; $F_{1, 6}=.31$, $p=.59$).
Older adults with dementia had no statistically significant improvement. However, their average score improved and less disperse.

The average score of emotional engagement increased by 4.49\%
(pre-test: $\Bar{M}=84.76$, $\sigma=14.25$; 
post-test: $\Bar{M}=88.57$, $\sigma=11.36$; $F_{1, 6}=.31$, $p=.59$). 
Older adults with dementia had no statistically significant improvement. However, their average score improved and less disperse.

The average score of cognitive engagement slightly decreased by 5.33\%
(pre-test: $\Bar{M}=71.43$, $\sigma=18.74$; 
post-test: $\Bar{M}=67.62$, $\sigma=17.82$; $F_{1, 6}=.15$, $p=.70$).

The average score of intrinsic motivation increased by 6.45\%
(pre-test: $\Bar{M}=88.57$, $\sigma=10.69$; 
post-test: $\Bar{M}=94.29$, $\sigma=9.76$; $F_{1, 6}=1.09$, $p=.32$).
Older adults with dementia had no statistically significant improvement. However, their average score improved and less disperse.

\subsection{RQ2: How does the usability of interactive robot-assisted scenario learning system impact user satisfaction of older adult with dementia?}

The mean SUS score for the \emph{TrainBo} system was 65.36 ($\sigma=16.17$), indicating ``acceptable" usability. Among seven participants, two of them rated 80 points or higher, three of them rated between 65 to 75 points, the rest rated 40 to 48 points.
Most participants told us that they liked \emph{TrainBo} system as the system can read them the question. E7 mentioned, ``I do not remember the word. The system reads me the questions, I can listen and understand what to do." 
Almost participants appreciated the robot as TrainBo is cute with its humanoid appearance, such as shaking hands. Incorporating with animations and sound effects, TrainBo makes the older adults with dementia feel happy and encouragement. E9 told us, ``The robot can speak to me with animations. It  makes me feel happier and has a sense of wisdom."
Some participants indicated that they prefer colourful design, especially in red colour as it is more sensitive and the association in Chinese culture. For the font style, such as font type and font size, also a key concern to them due to the presbyopia issues.

\subsection {Observation}
Our observation covered different perspectives, including colour preferences, understanding of the equipment, patience and engagement, need for clear explanations, communication challenges, enjoyment and engagement, visual aids, and relatability to daily life. We elaborated in details: 

\begin{enumerate}
    \item Colour Preferences: The older adults expressed a preference for red over blue, indicating that the design elements of the robot and the training materials should be personalized to align with the individual preferences of different participants. Red colour is associated with life-generating energy, celebrations and prosperity in the Chinsese culture, symbolizing luck, joy, and happiness.

    \item Understanding the Equipment: Some participants had difficulty identifying the recording microphone, suggesting a need for clearer instructions or visual cues to enhance understanding.

    \item Patience and Engagement: The older adults demonstrated remarkable patience throughout the one-hour sessions, showing no signs of frustration or agitation. We inquired about their fatigue levels, and they reported feeling comfortable and engaged during the activities.

    \item Need for Clear Explanations: Participants required clear and thorough explanations to grasp the tasks effectively. This highlights the importance of using straightforward language and providing ample context during training.

    \item Communication Challenges: One participant exhibited difficulty in verbal communication due to drooling, which may have affected their willingness to speak. This suggests that alternative communication methods or additional support may be necessary for some individuals.

    \item Enjoyment and Engagement: Many participants expressed enjoyment, stating they found the activities fun and appreciated the opportunity to learn new scenarios. They particularly enjoyed following the robot's actions, such as clapping along.

    \item Visual Aids: The use of images was noted to be effective, as participants found them easy to understand. When the robot provided praise for their efforts, participants responded positively, often expressing gratitude with comments like, "Thank you for the compliment."

    \item Relatability to Daily Life: Activities that closely resembled everyday situations were easier for participants to comprehend and engage with. For instance, when the robot sang and performed gestures, the older adults appeared particularly happy and involved.
\end{enumerate}

\section{Discussion} \label{sec:discussion}
In this section, we discuss the unique digital challenges that older adults with dementia faces, and how these challenges could be better addressed by the robot-assisted scenario training system.

\paragraph{Accessibility in Learning Design:} Challenges related to the use of technology and the navigation of the interface underscore the need for user-friendly design and adequate technical support to address unfamiliarity and confusion problems. Participants emphasize the importance of auditory learning and user-friendly interfaces to address age-related challenges. E8 mentioned `` I don't know how to read, I prefer visual elements" and E10 ``You read it aloud to me"; E6 told us ``Everyone want to hear the questions"; E2 remarked, ``I'm not very familiar with characters". They reflected the preference for instructions read aloud and struggles with unfamiliar technology.

\paragraph{Feedback and Encouragement: } Older adults with dementia value positive reinforcement, instructions, and feedback as essential motivators. Creating a supportive and encouraging learning environment can fosters feelings of competence and reduces fears of failure.  E1 shared ``I feel happy when answering correctly" and E6 said ``I want more encouragement". E7 also agreed ``Please give encourage to me". For their training exercise, seeking verbal praise, visual cues, and animated responses can enhance their sense of competence and mastery.

\paragraph{Personalized Learning Experiences:} Creating a more personalized and engaging learning journey can increase motivation and autonomy. E4 mentioned ``I want to choose the task by myself, like what I am doing in everyday"; E6 added ``I decide by myself"; E11 also shared the same feeling``I want to do this exercise". Therefore, the system should offer options for users to control their learning pace, choose feedback methods, and select activities based on their interests and daily scenarios.

\paragraph{}
While this study provides valuable insights, the small sample size limits and the short duration of the intervention may not fully capture the long-term effects of the TrainBo system on cognitive engagement and motivation. Besides, the limited diversity in the participant demographic, with a majority being female and older adults with varying levels of cognitive impairment, may also affect the findings. In addition, the language proficiency of the participant may limited their interest in interacting with the training system, thus affecting the effectiveness of the cognitive training.  

\section{Conclusion and Future Works} \label{sec:conclusion}

TrainBo demonstrates the promising role of interactive robot-assisted scenario training systems in enhancing engagement and motivation among older adults with dementia. By leveraging the principles of Self-Determination Theory, the system has shown potential in fostering intrinsic motivation and improving user satisfaction. The positive feedback from participants regarding the usability and interactive features of TrainBo suggests that such technological interventions can be valuable tools in dementia care. 

Future research could focus on several key areas to build upon the findings of this study. First, longitudinal studies could help to understand the long-term impacts of the TrainBo system on cognitive function, emotional well-being, and quality of life for older adults with dementia and investigating the scalability of such systems in diverse settings. Second, exploring the integration of advanced features, such as adaptive learning algorithms and personalized content to further enhance the effectiveness of the TrainBo system. Third, incorporating feedback mechanisms to allow TrainBo to adjust its interactions based on real-time user responses to increase engagement and motivation.

\bibliographystyle{ACM-Reference-Format}
\bibliography{sample-base}


\begin{thebibliography}{64}


\ifx \showCODEN    \undefined \def \showCODEN     #1{\unskip}     \fi
\ifx \showISBNx    \undefined \def \showISBNx     #1{\unskip}     \fi
\ifx \showISBNxiii \undefined \def \showISBNxiii  #1{\unskip}     \fi
\ifx \showISSN     \undefined \def \showISSN      #1{\unskip}     \fi
\ifx \showLCCN     \undefined \def \showLCCN      #1{\unskip}     \fi
\ifx \shownote     \undefined \def \shownote      #1{#1}          \fi
\ifx \showarticletitle \undefined \def \showarticletitle #1{#1}   \fi
\ifx \showURL      \undefined \def \showURL       {\relax}        \fi
\providecommand\bibfield[2]{#2}
\providecommand\bibinfo[2]{#2}
\providecommand\natexlab[1]{#1}
\providecommand\showeprint[2][]{arXiv:#2}

\bibitem[Appel et~al\mbox{.}(2021)]%
        {Appel2021VR}
\bibfield{author}{\bibinfo{person}{Laura Appel} {et~al\mbox{.}}} \bibinfo{year}{2021}\natexlab{}.
\newblock \showarticletitle{Virtual reality to promote wellbeing in persons with dementia: A scoping review}.
\newblock \bibinfo{journal}{\emph{Journal of Rehabilitation and Assistive Technologies Engineering}}  \bibinfo{volume}{8} (\bibinfo{year}{2021}).
\newblock
\href{https://doi.org/10.1177/20556683211053952}{doi:\nolinkurl{10.1177/20556683211053952}}


\bibitem[Arvanitakis et~al\mbox{.}(2019)]%
        {Arvanitakis2019}
\bibfield{author}{\bibinfo{person}{Zoe Arvanitakis}, \bibinfo{person}{Raj~C. Shah}, {and} \bibinfo{person}{David~A. Bennett}.} \bibinfo{year}{2019}\natexlab{}.
\newblock \showarticletitle{Diagnosis and Management of Dementia: Review}.
\newblock \bibinfo{journal}{\emph{JAMA}} \bibinfo{volume}{322}, \bibinfo{number}{16} (\bibinfo{date}{22 Oct.} \bibinfo{year}{2019}), \bibinfo{pages}{1589--1599}.
\newblock
\href{https://doi.org/10.1001/jama.2019.4782}{doi:\nolinkurl{10.1001/jama.2019.4782}}


\bibitem[Banducci et~al\mbox{.}(2017)]%
        {banducci2017active}
\bibfield{author}{\bibinfo{person}{Sarah~E Banducci}, \bibinfo{person}{Ana~M Daugherty}, \bibinfo{person}{John~R Biggan}, \bibinfo{person}{Gillian~E Cooke}, \bibinfo{person}{Michelle Voss}, \bibinfo{person}{Tony Noice}, \bibinfo{person}{Helga Noice}, {and} \bibinfo{person}{Arthur~F Kramer}.} \bibinfo{year}{2017}\natexlab{}.
\newblock \showarticletitle{Active experiencing training improves episodic memory recall in older adults}.
\newblock \bibinfo{journal}{\emph{Frontiers in aging neuroscience}}  \bibinfo{volume}{9} (\bibinfo{year}{2017}), \bibinfo{pages}{133}.
\newblock


\bibitem[Bangor et~al\mbox{.}(2008)]%
        {bangor2008empirical}
\bibfield{author}{\bibinfo{person}{Aaron Bangor}, \bibinfo{person}{Philip~T Kortum}, {and} \bibinfo{person}{James~T Miller}.} \bibinfo{year}{2008}\natexlab{}.
\newblock \showarticletitle{An empirical evaluation of the system usability scale}.
\newblock \bibinfo{journal}{\emph{Intl. Journal of Human--Computer Interaction}} \bibinfo{volume}{24}, \bibinfo{number}{6} (\bibinfo{year}{2008}), \bibinfo{pages}{574--594}.
\newblock


\bibitem[Bar-Tur(2021)]%
        {bar2021fostering}
\bibfield{author}{\bibinfo{person}{Liora Bar-Tur}.} \bibinfo{year}{2021}\natexlab{}.
\newblock \showarticletitle{Fostering well-being in the elderly: Translating theories on positive aging to practical approaches}.
\newblock \bibinfo{journal}{\emph{Frontiers in Medicine}}  \bibinfo{volume}{8} (\bibinfo{year}{2021}), \bibinfo{pages}{517226}.
\newblock


\bibitem[Boekaerts(1997)]%
        {boekaerts1997self}
\bibfield{author}{\bibinfo{person}{Monique Boekaerts}.} \bibinfo{year}{1997}\natexlab{}.
\newblock \showarticletitle{Self-regulated learning: A new concept embraced by researchers, policy makers, educators, teachers, and students}.
\newblock \bibinfo{journal}{\emph{Learning and instruction}} \bibinfo{volume}{7}, \bibinfo{number}{2} (\bibinfo{year}{1997}), \bibinfo{pages}{161--186}.
\newblock


\bibitem[Braun and Clarke(2006)]%
        {braun2006using}
\bibfield{author}{\bibinfo{person}{Virginia Braun} {and} \bibinfo{person}{Victoria Clarke}.} \bibinfo{year}{2006}\natexlab{}.
\newblock \showarticletitle{Using thematic analysis in psychology}.
\newblock \bibinfo{journal}{\emph{Qualitative research in psychology}} \bibinfo{volume}{3}, \bibinfo{number}{2} (\bibinfo{year}{2006}), \bibinfo{pages}{77--101}.
\newblock


\bibitem[Buchanan et~al\mbox{.}(2011)]%
        {buchanan2011role}
\bibfield{author}{\bibinfo{person}{Jeffrey~A Buchanan} {et~al\mbox{.}}} \bibinfo{year}{2011}\natexlab{}.
\newblock \showarticletitle{The role of behavior analysis in the rehabilitation of persons with dementia}.
\newblock \bibinfo{journal}{\emph{Behavior therapy}} \bibinfo{volume}{42}, \bibinfo{number}{1} (\bibinfo{year}{2011}), \bibinfo{pages}{9--21}.
\newblock


\bibitem[Cansev et~al\mbox{.}(2021)]%
        {cansev2021interactive}
\bibfield{author}{\bibinfo{person}{Mehmet~Ege Cansev} {et~al\mbox{.}}} \bibinfo{year}{2021}\natexlab{}.
\newblock \showarticletitle{Interactive human--robot skill transfer: a review of learning methods and user experience}.
\newblock \bibinfo{journal}{\emph{Advanced Intelligent Systems}} \bibinfo{volume}{3}, \bibinfo{number}{7} (\bibinfo{year}{2021}), \bibinfo{pages}{2000247}.
\newblock


\bibitem[Chang et~al\mbox{.}(2022)]%
        {Chang2022Interactive}
\bibfield{author}{\bibinfo{person}{Chia-Hao Chang} {et~al\mbox{.}}} \bibinfo{year}{2022}\natexlab{}.
\newblock \showarticletitle{Interactive Somatosensory Games in Rehabilitation training for Older adults with mild Cognitive Impairment: Usability Study}.
\newblock \bibinfo{journal}{\emph{JMIR Serious Games}} \bibinfo{volume}{10}, \bibinfo{number}{3} (\bibinfo{year}{2022}), \bibinfo{pages}{e38465}.
\newblock
\href{https://doi.org/10.2196/38465}{doi:\nolinkurl{10.2196/38465}}


\bibitem[Chang and Bourgeois(2020)]%
        {chang2020effects}
\bibfield{author}{\bibinfo{person}{Wan-Zu~D Chang} {and} \bibinfo{person}{Michelle~S Bourgeois}.} \bibinfo{year}{2020}\natexlab{}.
\newblock \showarticletitle{Effects of visual aids for end-of-life care on decisional capacity of people with dementia}.
\newblock \bibinfo{journal}{\emph{American Journal of speech-language Pathology}} \bibinfo{volume}{29}, \bibinfo{number}{1} (\bibinfo{year}{2020}), \bibinfo{pages}{185--200}.
\newblock


\bibitem[Cipriani et~al\mbox{.}(2013)]%
        {cipriani2013repetitive}
\bibfield{author}{\bibinfo{person}{Gabriele Cipriani} {et~al\mbox{.}}} \bibinfo{year}{2013}\natexlab{}.
\newblock \showarticletitle{Repetitive and stereotypic phenomena and dementia}.
\newblock \bibinfo{journal}{\emph{American Journal of Alzheimer's Disease \& Other Dementias{\textregistered}}} \bibinfo{volume}{28}, \bibinfo{number}{3} (\bibinfo{year}{2013}), \bibinfo{pages}{223--227}.
\newblock


\bibitem[Cipriani et~al\mbox{.}(2020)]%
        {cipriani2020daily}
\bibfield{author}{\bibinfo{person}{Gabriele Cipriani} {et~al\mbox{.}}} \bibinfo{year}{2020}\natexlab{}.
\newblock \showarticletitle{Daily functioning and dementia}.
\newblock \bibinfo{journal}{\emph{Dementia \& neuropsychologia}} \bibinfo{volume}{14}, \bibinfo{number}{2} (\bibinfo{year}{2020}), \bibinfo{pages}{93--102}.
\newblock


\bibitem[Coronado et~al\mbox{.}(2022)]%
        {coronado2022evaluating}
\bibfield{author}{\bibinfo{person}{Enrique Coronado} {et~al\mbox{.}}} \bibinfo{year}{2022}\natexlab{}.
\newblock \showarticletitle{Evaluating quality in human-robot interaction: {A} systematic search and classification of performance and human-centered factors, measures and metrics towards an industry 5.0}.
\newblock \bibinfo{journal}{\emph{Journal of Manufacturing Systems}}  \bibinfo{volume}{63} (\bibinfo{year}{2022}), \bibinfo{pages}{392--410}.
\newblock


\bibitem[Das et~al\mbox{.}(2024)]%
        {das2024chapter}
\bibfield{author}{\bibinfo{person}{Kiran~Sourav Das} {et~al\mbox{.}}} \bibinfo{year}{2024}\natexlab{}.
\newblock \showarticletitle{Chapter-9 Overview of Extension Teaching Methods}.
\newblock \bibinfo{journal}{\emph{A Comprehensive Textbook of Extension Education and Communication Management}} (\bibinfo{year}{2024}), \bibinfo{pages}{157}.
\newblock


\bibitem[Dewar et~al\mbox{.}(2014)]%
        {dewar2014boosting}
\bibfield{author}{\bibinfo{person}{Michaela Dewar} {et~al\mbox{.}}} \bibinfo{year}{2014}\natexlab{}.
\newblock \showarticletitle{Boosting long-term memory via wakeful rest: intentional rehearsal is not necessary, consolidation is sufficient}.
\newblock \bibinfo{journal}{\emph{PLoS one}} \bibinfo{volume}{9}, \bibinfo{number}{10} (\bibinfo{year}{2014}), \bibinfo{pages}{e109542}.
\newblock


\bibitem[Emmady et~al\mbox{.}(2025)]%
        {Emmady2025}
\bibfield{author}{\bibinfo{person}{Pradeep~D. Emmady}, \bibinfo{person}{Christopher Schoo}, {and} \bibinfo{person}{Prashant Tadi}.} \bibinfo{year}{2025}\natexlab{}.
\newblock \bibinfo{booktitle}{\emph{Major Neurocognitive Disorder (Dementia)}}.
\newblock Treasure Island, FL.
\newblock
\newblock
\shownote{[Updated 2022 Nov 19]}.


\bibitem[Fasola and Mataric(2012)]%
        {fasola2012using}
\bibfield{author}{\bibinfo{person}{Juan Fasola} {and} \bibinfo{person}{Maja~J Mataric}.} \bibinfo{year}{2012}\natexlab{}.
\newblock \showarticletitle{Using socially assistive human--robot interaction to motivate physical exercise for older adults}.
\newblock \bibinfo{journal}{\emph{Proc. IEEE}} \bibinfo{volume}{100}, \bibinfo{number}{8} (\bibinfo{year}{2012}), \bibinfo{pages}{2512--2526}.
\newblock


\bibitem[Forkan et~al\mbox{.}(2019)]%
        {Forkan2019}
\bibfield{author}{\bibinfo{person}{A.~R.~M. Forkan}, \bibinfo{person}{Peter Branch}, \bibinfo{person}{P.~P. Jayaraman}, {and} \bibinfo{person}{A. Ferretto}.} \bibinfo{year}{2019}\natexlab{}.
\newblock \showarticletitle{An Internet-of-Things Solution to Assist Independent Living and Social Connectedness in Elderly}.
\newblock \bibinfo{journal}{\emph{ACM Transactions on Social Computing}} \bibinfo{volume}{2}, \bibinfo{number}{4} (\bibinfo{year}{2019}), \bibinfo{pages}{1--24}.
\newblock
\href{https://doi.org/10.1145/3363563}{doi:\nolinkurl{10.1145/3363563}}


\bibitem[Fury(2024)]%
        {fury2024active}
\bibfield{author}{\bibinfo{person}{Sam Fury}.} \bibinfo{year}{2024}\natexlab{}.
\newblock \bibinfo{booktitle}{\emph{Active Mind Maintenance: Tools and Tips for Improving Cognitive Thinking}}.
\newblock \bibinfo{publisher}{SF Nonfiction Books}.
\newblock


\bibitem[Gagn{\'e} et~al\mbox{.}(2022)]%
        {Gagne2022UAS}
\bibfield{author}{\bibinfo{person}{Maryl{\`e}ne Gagn{\'e}} {et~al\mbox{.}}} \bibinfo{year}{2022}\natexlab{}.
\newblock \showarticletitle{Understanding and shaping the future of work with self-determination theory}.
\newblock \bibinfo{journal}{\emph{Nature Reviews Psychology}} \bibinfo{volume}{1}, \bibinfo{number}{7} (\bibinfo{year}{2022}), \bibinfo{pages}{378--392}.
\newblock
\href{https://doi.org/10.1038/s44159-022-00056-w}{doi:\nolinkurl{10.1038/s44159-022-00056-w}}


\bibitem[Gasteiger et~al\mbox{.}(2021)]%
        {Gasteiger2021Robot}
\bibfield{author}{\bibinfo{person}{Nina Gasteiger} {et~al\mbox{.}}} \bibinfo{year}{2021}\natexlab{}.
\newblock \showarticletitle{Robot-Delivered Cognitive Stimulation Games for Older Adults: Usability and Acceptability Evaluation}.
\newblock \bibinfo{journal}{\emph{ACM Transactions on Human-Robotic Interaction}} \bibinfo{volume}{10}, \bibinfo{number}{4} (\bibinfo{year}{2021}), \bibinfo{pages}{1--18}.
\newblock
\href{https://doi.org/10.1145/3451882}{doi:\nolinkurl{10.1145/3451882}}


\bibitem[Gongor and Tutsoy(2025)]%
        {gongor2025remarkable}
\bibfield{author}{\bibinfo{person}{Fatma Gongor} {and} \bibinfo{person}{Onder Tutsoy}.} \bibinfo{year}{2025}\natexlab{}.
\newblock \showarticletitle{On the remarkable advancement of assistive robotics in human-robot interaction-based health-care applications: An exploratory overview of the literature}.
\newblock \bibinfo{journal}{\emph{International Journal of Human--Computer Interaction}} \bibinfo{volume}{41}, \bibinfo{number}{2} (\bibinfo{year}{2025}), \bibinfo{pages}{1502--1542}.
\newblock


\bibitem[Goodall et~al\mbox{.}(2021)]%
        {goodall2021use}
\bibfield{author}{\bibinfo{person}{Gemma Goodall} {et~al\mbox{.}}} \bibinfo{year}{2021}\natexlab{}.
\newblock \showarticletitle{The use of technology in creating individualized, meaningful activities for people living with dementia: {A} systematic review}.
\newblock \bibinfo{journal}{\emph{Dementia}} \bibinfo{volume}{20}, \bibinfo{number}{4} (\bibinfo{year}{2021}), \bibinfo{pages}{1442--1469}.
\newblock


\bibitem[Guthrie et~al\mbox{.}(2018)]%
        {guthrie2018combined}
\bibfield{author}{\bibinfo{person}{Dawn~M Guthrie}, \bibinfo{person}{Jacob~GS Davidson}, \bibinfo{person}{Nicole Williams}, \bibinfo{person}{Jennifer Campos}, \bibinfo{person}{Kathleen Hunter}, \bibinfo{person}{Paul Mick}, \bibinfo{person}{Joseph~B Orange}, \bibinfo{person}{M~Kathleen Pichora-Fuller}, \bibinfo{person}{Natalie~A Phillips}, \bibinfo{person}{Marie~Y Savundranayagam}, {et~al\mbox{.}}} \bibinfo{year}{2018}\natexlab{}.
\newblock \showarticletitle{Combined impairments in vision, hearing and cognition are associated with greater levels of functional and communication difficulties than cognitive impairment alone: Analysis of interRAI data for home care and long-term care recipients in Ontario}.
\newblock \bibinfo{journal}{\emph{PLoS one}} \bibinfo{volume}{13}, \bibinfo{number}{2} (\bibinfo{year}{2018}), \bibinfo{pages}{e0192971}.
\newblock


\bibitem[Han et~al\mbox{.}(2024)]%
        {han2024human}
\bibfield{author}{\bibinfo{person}{In~Ho Han} {et~al\mbox{.}}} \bibinfo{year}{2024}\natexlab{}.
\newblock \showarticletitle{Human-Robot Interaction and Social Robot: The Emerging Field of Healthcare Robotics and Current and Future Perspectives for Spinal Care}.
\newblock \bibinfo{journal}{\emph{Neurospine}} \bibinfo{volume}{21}, \bibinfo{number}{3} (\bibinfo{year}{2024}), \bibinfo{pages}{868}.
\newblock


\bibitem[Hung et~al\mbox{.}(2019)]%
        {Hung2019Benefits}
\bibfield{author}{\bibinfo{person}{Linda Hung} {et~al\mbox{.}}} \bibinfo{year}{2019}\natexlab{}.
\newblock \showarticletitle{The benefits of and barriers to using a social robot PARO in care settings: a scoping review}.
\newblock \bibinfo{journal}{\emph{BMC Geriatrics}} \bibinfo{volume}{19}, \bibinfo{number}{1} (\bibinfo{year}{2019}).
\newblock
\href{https://doi.org/10.1186/s12877-019-1244-6}{doi:\nolinkurl{10.1186/s12877-019-1244-6}}


\bibitem[Kortum and Peres(2014)]%
        {kortum2014relationship}
\bibfield{author}{\bibinfo{person}{Philip Kortum} {and} \bibinfo{person}{S~Camille Peres}.} \bibinfo{year}{2014}\natexlab{}.
\newblock \showarticletitle{The relationship between system effectiveness and subjective usability scores using the System Usability Scale}.
\newblock \bibinfo{journal}{\emph{International Journal of Human-Computer Interaction}} \bibinfo{volume}{30}, \bibinfo{number}{7} (\bibinfo{year}{2014}), \bibinfo{pages}{575--584}.
\newblock


\bibitem[Kume et~al\mbox{.}(2023)]%
        {kume2023effect}
\bibfield{author}{\bibinfo{person}{Yu Kume} {et~al\mbox{.}}} \bibinfo{year}{2023}\natexlab{}.
\newblock \showarticletitle{Effect of a multicomponent programme based on reality orientation therapy on the physical performance and cognitive function of elderly community-dwellers: a quasi-experimental study}.
\newblock \bibinfo{journal}{\emph{Psychogeriatrics}} \bibinfo{volume}{23}, \bibinfo{number}{5} (\bibinfo{year}{2023}), \bibinfo{pages}{847--855}.
\newblock


\bibitem[Kurazume et~al\mbox{.}(2022)]%
        {kurazume2022development}
\bibfield{author}{\bibinfo{person}{Ryo Kurazume} {et~al\mbox{.}}} \bibinfo{year}{2022}\natexlab{}.
\newblock \showarticletitle{Development of AR training systems for Humanitude dementia care}.
\newblock \bibinfo{journal}{\emph{Advanced Robotics}} \bibinfo{volume}{36}, \bibinfo{number}{7} (\bibinfo{year}{2022}), \bibinfo{pages}{344--358}.
\newblock


\bibitem[Lewis(2018)]%
        {lewis2018system}
\bibfield{author}{\bibinfo{person}{James~R Lewis}.} \bibinfo{year}{2018}\natexlab{}.
\newblock \showarticletitle{The system usability scale: past, present, and future}.
\newblock \bibinfo{journal}{\emph{International Journal of Human--Computer Interaction}} \bibinfo{volume}{34}, \bibinfo{number}{7} (\bibinfo{year}{2018}), \bibinfo{pages}{577--590}.
\newblock


\bibitem[Liu et~al\mbox{.}(2023)]%
        {liu2023older}
\bibfield{author}{\bibinfo{person}{Mingzhou Liu} {et~al\mbox{.}}} \bibinfo{year}{2023}\natexlab{}.
\newblock \showarticletitle{Older adults’ intention to use voice assistants: Usability and emotional needs}.
\newblock \bibinfo{journal}{\emph{Heliyon}} \bibinfo{volume}{9}, \bibinfo{number}{11} (\bibinfo{year}{2023}).
\newblock


\bibitem[Liverman et~al\mbox{.}(2015)]%
        {liverman2015cognitive}
\bibfield{author}{\bibinfo{person}{Catharyn~T Liverman}, \bibinfo{person}{Kristine Yaffe}, {and} \bibinfo{person}{Dan~G Blazer}.} \bibinfo{year}{2015}\natexlab{}.
\newblock \showarticletitle{Cognitive aging: Progress in understanding and opportunities for action}.
\newblock  (\bibinfo{year}{2015}).
\newblock


\bibitem[Lobbia et~al\mbox{.}(2018)]%
        {Lobbia2018Cognitive}
\bibfield{author}{\bibinfo{person}{Anna Lobbia} {et~al\mbox{.}}} \bibinfo{year}{2018}\natexlab{}.
\newblock \showarticletitle{Cognitive stimulation therapy in dementia: {A} systematic review of neuroplasticity and efficacy}.
\newblock \bibinfo{journal}{\emph{Neuropsychological Rehabilitation}} \bibinfo{volume}{28}, \bibinfo{number}{5} (\bibinfo{year}{2018}), \bibinfo{pages}{714--733}.
\newblock
\href{https://doi.org/10.1080/09602011.2017.1357612}{doi:\nolinkurl{10.1080/09602011.2017.1357612}}


\bibitem[Man et~al\mbox{.}(2012)]%
        {man2012evaluation}
\bibfield{author}{\bibinfo{person}{David~WK Man}, \bibinfo{person}{Jenny~CC Chung}, {and} \bibinfo{person}{Grace~YY Lee}.} \bibinfo{year}{2012}\natexlab{}.
\newblock \showarticletitle{Evaluation of a virtual reality-based memory training programme for Hong Kong Chinese older adults with questionable dementia: a pilot study}.
\newblock \bibinfo{journal}{\emph{International journal of geriatric psychiatry}} \bibinfo{volume}{27}, \bibinfo{number}{5} (\bibinfo{year}{2012}), \bibinfo{pages}{513--520}.
\newblock


\bibitem[Mezrar and Bendella(2022)]%
        {Mezrar2022ASR}
\bibfield{author}{\bibinfo{person}{Samiha Mezrar} {and} \bibinfo{person}{Fatima Bendella}.} \bibinfo{year}{2022}\natexlab{}.
\newblock \showarticletitle{{A} Systematic Review of Serious Games Relating to Cognitive Impairment and Dementia}.
\newblock \bibinfo{journal}{\emph{Journal of Digital Information Management}} \bibinfo{volume}{20}, \bibinfo{number}{1} (\bibinfo{year}{2022}), \bibinfo{pages}{1--9}.
\newblock


\bibitem[Mor{\'a}n et~al\mbox{.}(2024)]%
        {moran2024serious}
\bibfield{author}{\bibinfo{person}{Juan Francisco~Ortega Mor{\'a}n}, \bibinfo{person}{J~Blas Pagador}, \bibinfo{person}{Vicente~Gilete Preciado}, \bibinfo{person}{Jos{\'e}~Luis Moyano-Cuevas}, \bibinfo{person}{Trinidad~Rodr{\'\i}guez Dom{\'\i}nguez}, \bibinfo{person}{Marta~Santurino Mu{\~n}oz}, {and} \bibinfo{person}{Francisco M~S{\'a}nchez Margallo}.} \bibinfo{year}{2024}\natexlab{}.
\newblock \showarticletitle{A serious game for cognitive stimulation of older people with mild cognitive impairment: Design and pilot usability study}.
\newblock \bibinfo{journal}{\emph{JMIR aging}} \bibinfo{volume}{7}, \bibinfo{number}{1} (\bibinfo{year}{2024}), \bibinfo{pages}{e41437}.
\newblock


\bibitem[Moro et~al\mbox{.}(2019)]%
        {moro2019}
\bibfield{author}{\bibinfo{person}{C. Moro}, \bibinfo{person}{S. Lin}, \bibinfo{person}{G. Nejat}, {and} \bibinfo{person}{A. Mihailidis}.} \bibinfo{year}{2019}\natexlab{}.
\newblock \showarticletitle{Social robots and seniors: A comparative study on the influence of dynamic social features on human–robot interaction}.
\newblock \bibinfo{journal}{\emph{International Journal of Social Robotics}}  \bibinfo{volume}{11} (\bibinfo{year}{2019}), \bibinfo{pages}{5--24}.
\newblock
\href{https://doi.org/10.1007/s12369-018-0488-1}{doi:\nolinkurl{10.1007/s12369-018-0488-1}}


\bibitem[Moyle et~al\mbox{.}(2022)]%
        {Moyle2022Therapeutic}
\bibfield{author}{\bibinfo{person}{Wendy Moyle} {et~al\mbox{.}}} \bibinfo{year}{2022}\natexlab{}.
\newblock \showarticletitle{Therapeutic use of the humanoid robot, Telenoid, with older adults: A critical interpretive synthesis review}.
\newblock \bibinfo{journal}{\emph{Assistive Technology}} \bibinfo{volume}{36}, \bibinfo{number}{5} (\bibinfo{year}{2022}), \bibinfo{pages}{388--395}.
\newblock
\href{https://doi.org/10.1080/10400435.2022.2060375}{doi:\nolinkurl{10.1080/10400435.2022.2060375}}


\bibitem[Moyle et~al\mbox{.}(2020)]%
        {moyle2020me}
\bibfield{author}{\bibinfo{person}{Wendy Moyle}, \bibinfo{person}{Cindy Jones}, \bibinfo{person}{Jenny Murfield}, {and} \bibinfo{person}{Fangli Liu}.} \bibinfo{year}{2020}\natexlab{}.
\newblock \showarticletitle{‘{F}or me at 90, it’s going to be difficult’: feasibility of using i{P}ad video-conferencing with older adults in long-term aged care}.
\newblock \bibinfo{journal}{\emph{Aging \& mental health}} \bibinfo{volume}{24}, \bibinfo{number}{2} (\bibinfo{year}{2020}), \bibinfo{pages}{349--352}.
\newblock


\bibitem[Nault et~al\mbox{.}(2025b)]%
        {Nault2025Socially}
\bibfield{author}{\bibinfo{person}{Emeline Nault} {et~al\mbox{.}}} \bibinfo{year}{2025}\natexlab{b}.
\newblock \showarticletitle{Socially Assistive Robots and Sensory Feedback for Engaging Older Adults in Cognitive Activities}.
\newblock \bibinfo{journal}{\emph{ACM Transactions on Human-Robotic Interaction}} \bibinfo{volume}{14}, \bibinfo{number}{1} (\bibinfo{year}{2025}), \bibinfo{pages}{1--26}.
\newblock
\href{https://doi.org/10.1145/3698241}{doi:\nolinkurl{10.1145/3698241}}


\bibitem[Nault et~al\mbox{.}(2025a)]%
        {Nault2025}
\bibfield{author}{\bibinfo{person}{Emilyann Nault}, \bibinfo{person}{Lynne Baillie}, {and} \bibinfo{person}{Frank Broz}.} \bibinfo{year}{2025}\natexlab{a}.
\newblock \showarticletitle{Socially Assistive Robots and Sensory Feedback for Engaging Older Adults in Cognitive Activities}.
\newblock \bibinfo{journal}{\emph{ACM Transactions on Human-Robot Interaction}} \bibinfo{volume}{14}, \bibinfo{number}{1} (\bibinfo{year}{2025}), \bibinfo{pages}{1--26}.
\newblock
\href{https://doi.org/10.1145/3698241}{doi:\nolinkurl{10.1145/3698241}}


\bibitem[{NUWA Robotics Corp.}(2024)]%
        {nuwa2024kebbi}
\bibfield{author}{\bibinfo{person}{{NUWA Robotics Corp.}}} \bibinfo{year}{2024}\natexlab{}.
\newblock \bibinfo{title}{Kebbi Robots - Create a Personalized Robot with Voice Intelligence}.
\newblock \bibinfo{howpublished}{\url{https://www.nuwarobotics.com/en/product/}}.
\newblock
\newblock
\shownote{Accessed: 2025-04-18}.


\bibitem[Ostrowski et~al\mbox{.}(2019)]%
        {ostrowski2019older}
\bibfield{author}{\bibinfo{person}{Anastasia~K Ostrowski} {et~al\mbox{.}}} \bibinfo{year}{2019}\natexlab{}.
\newblock \showarticletitle{Older adults living with social robots: promoting social connectedness in long-term communities}.
\newblock \bibinfo{journal}{\emph{IEEE Robotics \& Automation Magazine}} \bibinfo{volume}{26}, \bibinfo{number}{2} (\bibinfo{year}{2019}), \bibinfo{pages}{59--70}.
\newblock


\bibitem[Paradowski and Jeli{\'n}ska(2024)]%
        {paradowski2024predictors}
\bibfield{author}{\bibinfo{person}{Micha{\l}~B Paradowski} {and} \bibinfo{person}{Magdalena Jeli{\'n}ska}.} \bibinfo{year}{2024}\natexlab{}.
\newblock \showarticletitle{The predictors of {L2} grit and their complex interactions in online foreign language learning: motivation, self-directed learning, autonomy, curiosity, and language mindsets}.
\newblock \bibinfo{journal}{\emph{Computer Assisted Language Learning}} \bibinfo{volume}{37}, \bibinfo{number}{8} (\bibinfo{year}{2024}), \bibinfo{pages}{2320--2358}.
\newblock


\bibitem[Paterson(2023)]%
        {paterson2023social}
\bibfield{author}{\bibinfo{person}{Mark Paterson}.} \bibinfo{year}{2023}\natexlab{}.
\newblock \showarticletitle{Social robots and the futures of affective touch}.
\newblock \bibinfo{journal}{\emph{The Senses and Society}} \bibinfo{volume}{18}, \bibinfo{number}{2} (\bibinfo{year}{2023}), \bibinfo{pages}{110--125}.
\newblock


\bibitem[Pirhonen et~al\mbox{.}(2020)]%
        {pirhonen2020could}
\bibfield{author}{\bibinfo{person}{Jari Pirhonen} {et~al\mbox{.}}} \bibinfo{year}{2020}\natexlab{}.
\newblock \showarticletitle{Could robots strengthen the sense of autonomy of older people residing in assisted living facilities?—A future-oriented study}.
\newblock \bibinfo{journal}{\emph{Ethics and Information Technology}} \bibinfo{volume}{22}, \bibinfo{number}{2} (\bibinfo{year}{2020}), \bibinfo{pages}{151--162}.
\newblock


\bibitem[Puebla et~al\mbox{.}(2022)]%
        {puebla2022mobile}
\bibfield{author}{\bibinfo{person}{Cecilia Puebla} {et~al\mbox{.}}} \bibinfo{year}{2022}\natexlab{}.
\newblock \showarticletitle{Mobile-assisted language learning in older adults: Chances and challenges}.
\newblock \bibinfo{journal}{\emph{ReCALL}} \bibinfo{volume}{34}, \bibinfo{number}{2} (\bibinfo{year}{2022}), \bibinfo{pages}{169--184}.
\newblock


\bibitem[Rai et~al\mbox{.}(2020)]%
        {rai2020individual}
\bibfield{author}{\bibinfo{person}{Harleen~Kaur Rai} {et~al\mbox{.}}} \bibinfo{year}{2020}\natexlab{}.
\newblock \showarticletitle{An individual cognitive stimulation therapy app for people with dementia: {D}evelopment and usability study of thinkability}.
\newblock \bibinfo{journal}{\emph{JMIR aging}} \bibinfo{volume}{3}, \bibinfo{number}{2} (\bibinfo{year}{2020}), \bibinfo{pages}{e17105}.
\newblock


\bibitem[Rai et~al\mbox{.}(2022)]%
        {Rai2022Digital}
\bibfield{author}{\bibinfo{person}{Hem~K Rai} {et~al\mbox{.}}} \bibinfo{year}{2022}\natexlab{}.
\newblock \showarticletitle{Digital Technologies to Prevent Social isolation and Loneliness in Dementia: {A} Systematic review}.
\newblock \bibinfo{journal}{\emph{Journal of Alzheimer's Disease}} \bibinfo{volume}{90}, \bibinfo{number}{2} (\bibinfo{year}{2022}), \bibinfo{pages}{513--528}.
\newblock
\href{https://doi.org/10.3233/JAD-220438}{doi:\nolinkurl{10.3233/JAD-220438}}


\bibitem[Razavi et~al\mbox{.}(2022)]%
        {Razavi2022}
\bibfield{author}{\bibinfo{person}{Seyedeh~Zahra Razavi}, \bibinfo{person}{Lenhart~K. Schubert}, \bibinfo{person}{Kimberly Van~Orden}, \bibinfo{person}{Mohammad~Rafayet Ali}, \bibinfo{person}{Benjamin Kane}, {and} \bibinfo{person}{Ehsan Hoque}.} \bibinfo{year}{2022}\natexlab{}.
\newblock \showarticletitle{Discourse Behavior of Older Adults Interacting with a Dialogue Agent Competent in Multiple Topics}.
\newblock \bibinfo{journal}{\emph{ACM Transactions on Interactive Intelligent Systems}} \bibinfo{volume}{12}, \bibinfo{number}{2} (\bibinfo{year}{2022}), \bibinfo{pages}{1--21}.
\newblock
\href{https://doi.org/10.1145/3484510}{doi:\nolinkurl{10.1145/3484510}}


\bibitem[Ryan and Deci(2017)]%
        {ryan2017self}
\bibfield{author}{\bibinfo{person}{Richard~M Ryan} {and} \bibinfo{person}{Edward~L Deci}.} \bibinfo{year}{2017}\natexlab{}.
\newblock \bibinfo{booktitle}{\emph{Self-determination theory: Basic psychological needs in motivation, development, and wellness}}.
\newblock \bibinfo{publisher}{Guilford publications}.
\newblock


\bibitem[Ryu et~al\mbox{.}(2020)]%
        {ryu2020simple}
\bibfield{author}{\bibinfo{person}{Hyeyoung Ryu}, \bibinfo{person}{Soyeon Kim}, \bibinfo{person}{Dain Kim}, \bibinfo{person}{Sooan Han}, \bibinfo{person}{Keeheon Lee}, {and} \bibinfo{person}{Younah Kang}.} \bibinfo{year}{2020}\natexlab{}.
\newblock \showarticletitle{Simple and steady interactions win the healthy mentality: designing a chatbot service for the elderly}.
\newblock \bibinfo{journal}{\emph{Proceedings of the ACM on human-computer interaction}} \bibinfo{volume}{4}, \bibinfo{number}{CSCW2} (\bibinfo{year}{2020}), \bibinfo{pages}{1--25}.
\newblock


\bibitem[Sachdev et~al\mbox{.}(2014)]%
        {Sachdev2014}
\bibfield{author}{\bibinfo{person}{Perminder~S. Sachdev}, \bibinfo{person}{Deborah Blacker}, \bibinfo{person}{Dan~G. Blazer}, \bibinfo{person}{Mary Ganguli}, \bibinfo{person}{Dilip~V. Jeste}, \bibinfo{person}{Jane~S. Paulsen}, {and} \bibinfo{person}{Ronald~C. Petersen}.} \bibinfo{year}{2014}\natexlab{}.
\newblock \showarticletitle{Classifying neurocognitive disorders: the DSM-5 approach}.
\newblock \bibinfo{journal}{\emph{Nature Reviews Neurology}} \bibinfo{volume}{10}, \bibinfo{number}{11} (\bibinfo{date}{Nov.} \bibinfo{year}{2014}), \bibinfo{pages}{634--642}.
\newblock
\href{https://doi.org/10.1038/nrneurol.2014.181}{doi:\nolinkurl{10.1038/nrneurol.2014.181}}


\bibitem[Samuelsson et~al\mbox{.}(2020)]%
        {Samuelsson2020}
\bibfield{author}{\bibinfo{person}{Christina Samuelsson}, \bibinfo{person}{Ulrika Ferm}, {and} \bibinfo{person}{Anna Ekström}.} \bibinfo{year}{2020}\natexlab{}.
\newblock \showarticletitle{"It’s Our Gang" - Promoting Social Inclusion for People with Dementia by Using Digital Communication Support in a Group Activity}.
\newblock \bibinfo{journal}{\emph{Clinical Gerontologist}} \bibinfo{volume}{44}, \bibinfo{number}{4} (\bibinfo{year}{2020}), \bibinfo{pages}{418--429}.
\newblock
\href{https://doi.org/10.1080/07317115.2020.1795037}{doi:\nolinkurl{10.1080/07317115.2020.1795037}}


\bibitem[Smith et~al\mbox{.}(2011)]%
        {smith2011memory}
\bibfield{author}{\bibinfo{person}{Erin~R Smith} {et~al\mbox{.}}} \bibinfo{year}{2011}\natexlab{}.
\newblock \showarticletitle{Memory and communication support in dementia: {R}esearch-based strategies for caregivers}.
\newblock \bibinfo{journal}{\emph{International {P}sychogeriatrics}} \bibinfo{volume}{23}, \bibinfo{number}{2} (\bibinfo{year}{2011}), \bibinfo{pages}{256--263}.
\newblock


\bibitem[Spector et~al\mbox{.}(2003)]%
        {Spector2003}
\bibfield{author}{\bibinfo{person}{A. Spector} {et~al\mbox{.}}} \bibinfo{year}{2003}\natexlab{}.
\newblock \showarticletitle{Efficacy of an evidence-based cognitive stimulation therapy programme for people with dementia: randomised controlled trial}.
\newblock \bibinfo{journal}{\emph{The British Journal of Psychiatry}} \bibinfo{volume}{183}, \bibinfo{number}{3} (\bibinfo{year}{2003}), \bibinfo{pages}{248--254}.
\newblock
\href{https://doi.org/10.1192/bjp.183.3.248}{doi:\nolinkurl{10.1192/bjp.183.3.248}}


\bibitem[Surr et~al\mbox{.}(2017)]%
        {surr2017effective}
\bibfield{author}{\bibinfo{person}{Claire~A Surr} {et~al\mbox{.}}} \bibinfo{year}{2017}\natexlab{}.
\newblock \showarticletitle{Effective dementia education and training for the health and social care workforce: {A} systematic review of the literature}.
\newblock \bibinfo{journal}{\emph{Review of Educational Research}} \bibinfo{volume}{87}, \bibinfo{number}{5} (\bibinfo{year}{2017}), \bibinfo{pages}{966--1002}.
\newblock


\bibitem[Trend(2023)]%
        {trend2023measuring}
\bibfield{author}{\bibinfo{person}{UIUX Trend}.} \bibinfo{year}{2023}\natexlab{}.
\newblock \bibinfo{title}{Measuring and interpreting system usability scale {(SUS)}}.
\newblock


\bibitem[Tsantali et~al\mbox{.}(2017)]%
        {Tsantali2017Cognitive}
\bibfield{author}{\bibinfo{person}{Evdokia Tsantali} {et~al\mbox{.}}} \bibinfo{year}{2017}\natexlab{}.
\newblock \showarticletitle{Cognitive training versus cognitive stimulation in mild dementia: A randomized controlled trial}.
\newblock \bibinfo{journal}{\emph{American Journal of Alzheimer's Disease \& Other Dementias}} \bibinfo{volume}{32}, \bibinfo{number}{5} (\bibinfo{year}{2017}), \bibinfo{pages}{283--291}.
\newblock
\href{https://doi.org/10.1177/1533317517711324}{doi:\nolinkurl{10.1177/1533317517711324}}


\bibitem[Udjaja et~al\mbox{.}(2021)]%
        {Udjaja2021Healthy}
\bibfield{author}{\bibinfo{person}{Yogi Udjaja} {et~al\mbox{.}}} \bibinfo{year}{2021}\natexlab{}.
\newblock \showarticletitle{Healthy Elder: Brain Stimulation Game for the Elderly to Reduce the Risk of Dementia}.
\newblock \bibinfo{journal}{\emph{Procedia Computer Science}}  \bibinfo{volume}{179} (\bibinfo{year}{2021}), \bibinfo{pages}{95--102}.
\newblock


\bibitem[Valenzuela and Sachdev(2006)]%
        {Valenzuela2006}
\bibfield{author}{\bibinfo{person}{Michael~J. Valenzuela} {and} \bibinfo{person}{Perminder~S. Sachdev}.} \bibinfo{year}{2006}\natexlab{}.
\newblock \showarticletitle{Brain reserve and dementia: {A} systematic review}.
\newblock \bibinfo{journal}{\emph{Psychol. Med.}} \bibinfo{volume}{36}, \bibinfo{number}{4} (\bibinfo{year}{2006}), \bibinfo{pages}{441--454}.
\newblock
\href{https://doi.org/10.1017/S0033291705007139}{doi:\nolinkurl{10.1017/S0033291705007139}}


\bibitem[Xue et~al\mbox{.}(2023)]%
        {Xue2023VRNPT}
\bibfield{author}{\bibinfo{person}{Cheng Xue} {et~al\mbox{.}}} \bibinfo{year}{2023}\natexlab{}.
\newblock \showarticletitle{VRNPT: a neuropsychological test tool for diagnosing mild cognitive impairment using virtual reality and {EEG} signals}.
\newblock \bibinfo{journal}{\emph{International Journal of Human-Computer Interaction}} \bibinfo{volume}{40}, \bibinfo{number}{20} (\bibinfo{year}{2023}), \bibinfo{pages}{6268--6286}.
\newblock
\href{https://doi.org/10.1080/10447318.2023.2250605}{doi:\nolinkurl{10.1080/10447318.2023.2250605}}


\bibitem[Yu et~al\mbox{.}(2022)]%
        {Yu2022Socially}
\bibfield{author}{\bibinfo{person}{Clare Yu} {et~al\mbox{.}}} \bibinfo{year}{2022}\natexlab{}.
\newblock \showarticletitle{Socially assistive robots for people with dementia: systematic review and meta-analysis of feasibility, acceptability and the effect on cognition, neuropsychiatric symptoms and quality of life}.
\newblock \bibinfo{journal}{\emph{Ageing Research Reviews}}  \bibinfo{volume}{78} (\bibinfo{year}{2022}), \bibinfo{pages}{101633}.
\newblock


\end{thebibliography}

\end{document}